\documentclass[aps,preprintnumbers,superscriptaddress,nofootinbib,aps,prd,floatfix]{revtex4-1}
\pdfoutput=1 \usepackage{graphicx}
\usepackage{amsthm} \usepackage{amsmath,amssymb,physics}
\usepackage{url} \usepackage{hyperref} \usepackage{xcolor}

\definecolor{nicered}{rgb}{0.5,0.,0.}
\definecolor{nicegreen}{rgb}{0.,0.5,0.}
\definecolor{niceblue}{rgb}{0.,0.,0.5}
\definecolor{darkpink}{rgb}{0.8,0.47,0.47}
\hypersetup{colorlinks,citecolor=nicegreen,linkcolor=nicered,urlcolor=niceblue}
\usepackage{enumitem} \usepackage[normalem]{ulem} \setlist{nolistsep}
 \usepackage{setspace} \newcommand{\GeV}{\textrm{GeV}}

\begin{document}
\preprint{MSUHEP-22-020, PITT-PACC-2210, SMU-HEP-22-03} \title{ Parton
  distributions need representative sampling} \author{Aurore Courtoy}
\email{aurore@fisica.unam.mx} \affiliation{Instituto de F\'isica,
  Universidad Nacional Aut\'onoma de M\'exico, Apartado Postal 20-364,
  01000 Ciudad de M\'exico, Mexico\looseness=-1}
\author{Joey Huston} \email{huston@msu.edu} \affiliation{Department of
  Physics and
  Astronomy, Michigan State University, East Lansing, MI 48824,
  USA\looseness=-1}
\author{Pavel Nadolsky} \email{nadolsky@smu.edu}
\affiliation{Department of Physics, Southern Methodist University,
  Dallas, TX 75275-0181, USA} \author{Keping Xie}
\email{xiekeping@pitt.edu} \affiliation{Pittsburgh Particle Physics,
  Astrophysics, and Cosmology Center, Department of Physics and
  Astronomy, University of Pittsburgh, Pittsburgh, PA 15260,
  USA\looseness=-1}
\author{Mengshi Yan} \email{msyan@pku.edu.cn} \affiliation{School of
  Physics and State Key Laboratory of Nuclear
  Physics and Technology, Peking University, Beijing 100871,
  China\looseness=-1}
\author{C.-P. Yuan} \email{yuanch@msu.edu} \affiliation{Department of
  Physics and Astronomy, Michigan State
  University, East Lansing, MI 48824, USA\looseness=-1}

 \date{February 16, 2023}

\begin{abstract}
In global QCD fits of parton distribution functions (PDFs), a large
part of the estimated uncertainty on the PDFs originates from the
choices of parametric functional forms and fitting methodology.  We
argue that these types of uncertainties can be underestimated with
common PDF ensembles in high-stake measurements at the Large Hadron
Collider and Tevatron. A fruitful approach to quantify these
uncertainties is to view them as arising from sampling of allowed PDF
solutions in a multidimensional parametric space. This approach
applies powerful insights gained in recent statistical studies of
large-scale population surveys and quasi-Monte Carlo integration
methods. In particular, PDF fits may be affected by the big data
paradox, which stipulates that more experimental data do not
automatically raise the accuracy of PDFs -- close attention to the
data quality and sampling of possible PDF solutions is as
essential. To test if the sampling of the PDF uncertainty of an
experimental observable is truly representative of all acceptable
solutions, we introduce a technique (``a hopscotch scan'') based on a
combination of parameter scans and stochastic sampling. With this
technique, we show that the PDF uncertainty on key LHC cross sections
at 13 TeV obtained with the public NNPDF4.0 fitting code is larger
than the nominal uncertainty obtained with the published NNPDF4.0
Monte-Carlo replica sets, when accounting for the likelihood
distribution. On the same grounds, the uncertainties on the charm
distribution at a large momentum fraction $x$ and gluon PDF at small
$x$ are enlarged.  In PDF ensembles obtained in the analytic
minimization (Hessian) formalism, the tolerance on the PDF uncertainty
must be based on sufficiently complete sampling of PDF functional
forms and choices of the experiments.
\\

\noindent This updated version includes detailed comparisons that are supplementary to the journal version in Phys.Rev.D 107 (2023) 3, 034008.
\end{abstract}

\maketitle
\clearpage \newpage

\section{Introduction}

Precision phenomenology at hadron colliders relies upon accurate
predictions in the Standard Model (SM). An overwhelming number of such
theoretical predictions require parton distribution functions (PDFs)
in a proton, the nonperturbative functions $f_a(x,Q)$ quantifying
probabilities for finding quarks and gluons in a proton at an energy
scale $Q$ above 1 GeV. Multiple groups
\cite{Abramowicz:2015mha,Accardi:2016qay,Alekhin:2017kpj,Ball:2017nwa,Hou:2019efy,Bailey:2019yze,Bailey:2020ooq,Ball:2021leu,ATLAS:2021vod}
provide increasingly sophisticated parametrizations of PDFs by fitting
a growing collection of precise experimental data sets to advanced
multiloop calculations. High-luminosity (HL) measurements at the Large
Hadron Collider (LHC) and planned DIS experiments (Electron-Ion
Collider~\cite{AbdulKhalek:2021gbh}, Large Hadron Electron
Collider~\cite{LHeC:2020van}, Muon-Ion Collider~\cite{Acosta:2021qpx}
\ldots), combined with the progress in perturbative QCD calculations,
open opportunities both to learn about the PDFs and to find their new
applications. The global QCD analysis to determine the PDFs can be now
attempted by a broad circle of users thanks to the publicly available
{\tt xFitter} \cite{Alekhin:2014irh} and {\tt NNPDF}
\cite{Ball:2021leu} fitting codes. A recent whitepaper
\cite{Amoroso:2022eow} contributed to the Snowmass'2021 Summer Study
reviews ongoing progress in the PDF analysis.

\begin{figure}[b]
\centering \includegraphics[
  height=162pt,keepaspectratio]{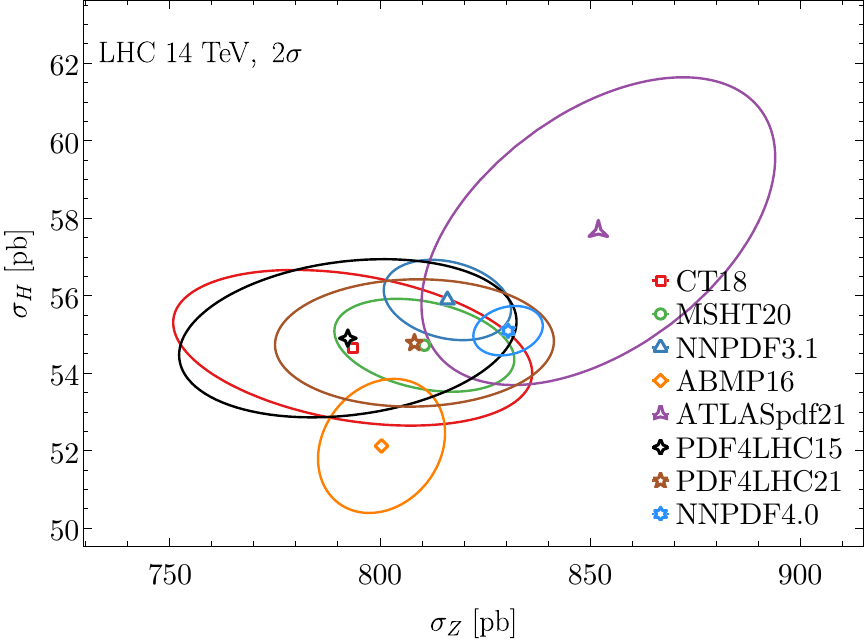}\quad\quad
\includegraphics[
  height=162pt,keepaspectratio]{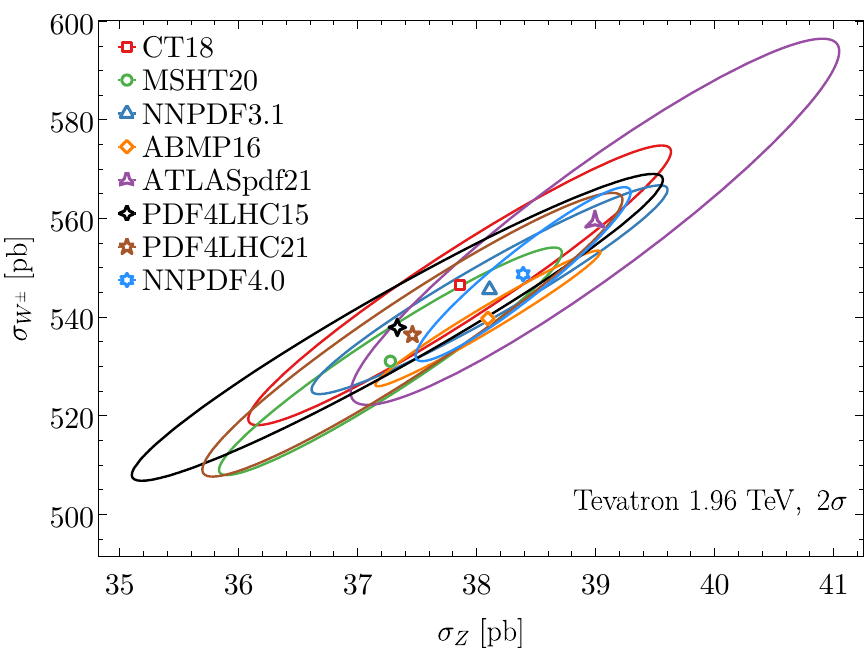}
\caption{NNLO theoretical predictions for $95\ \%$ C.L. PDF
  uncertainties for total cross sections of $Z$ and SM Higgs boson
  production at the LHC 14 TeV (left) and $Z$ and $W^{\pm}$ boson
  production at the Tevatron 1.96 TeV (right). The Higgs cross
  sections are obtained at NNLO multiplied by an N3LO/NNLO $K$ factor
  of 1.097 and by an EW $K$ factor of 1.0514. Predictions are shown
  for PDF4LHC21~\cite{PDF4LHCWorkingGroup:2022cjn},
  PDF4LHC15~\cite{Butterworth:2015oua}, NNPDF4.0~\cite{Ball:2021leu},
  CT18~\cite{Hou:2019efy}, MSHT20~\cite{Bailey:2020ooq},
  ABMP16~\cite{Alekhin:2017kpj}, and ATLASpdf21~\cite{ATLAS:2021vod}
  NNLO PDFs with $\alpha_s(M_Z)=0.118$.
 \label{fig:LHCTeVsigmaTot}}
\end{figure}

In this article, we summarize a study of a rarely discussed source of
some observed differences between the published parton
distributions. A lot of attention has been dedicated to various
factors that determine the accuracy of PDFs, usually associated with a
combination of experimental, theoretical, PDF parametrization, and
methodological sources. Estimates of PDF uncertainties are needed for
inference from QCD experiments at the LHC and other
facilities~\cite{Kovarik:2019xvh,PDF4LHCWorkingGroup:2022cjn,Gao:2017yyd}.
However, in addition to the accuracy of individual PDF fits, or
``fitting accuracy", another factor in the total uncertainty may be as
consequential, reflecting the accuracy of exploration, or {\it
  sampling}, of the space of acceptable PDF solutions. This space is
truly vast when large data samples are fitted using many
parameters. In fact, its exploration can be notoriously difficult, as
the sampling of multidimensional spaces is exponentially inefficient
\cite{Bellman:1961,Bishop:2006}.

In this context, sampled solutions can be obtained by varying the
fitted data (e.g., by resampling random fluctuations in the measured
values as in the NNPDF analyses \cite{Ball:2021leu}), models of theory
and experiment (e.g., the PDF parametrization forms \cite{Hou:2019efy}
or model parameters \cite{Watt:2012tq,Hou:2016sho} as in the CT and
MSHT studies), and in other ways. The pivotal role of adequate
sampling in large-scale data analyses has been emphasized in
statistics applications across diverse fields, notably in connection
with large-scale population surveys \cite{MengXL:2018,
  BradleyMeng:2021a}, multi-dimensional quasi-Monte Carlo (QMC)
integration \cite{Hickernell:2018a}, medical research
\cite{Msaouel:2022a}, variance-bias separation in machine learning
\cite{Geman:1992a,Bishop:2006}, and studies of predictivity of complex
models \cite{Puy:2022a}. Sampling issues are also pertinent to the
multivariate PDF fits. One key observation from the above studies is
that large samples do not guarantee convergence to the correct
solution, contrary to the common expectation based on the law of large
numbers. The reason is that nominally small biases in sampling of
possible solutions, such as in the selection of best-fit models of
PDFs obtained using various choices of experiments or functional forms
of PDFs, may {\it grow} as the volume of fitted data and complexity of
the analysis increase.  The growth reflects common difficulties with
sampling of parameter spaces of high dimensionality. In an
(unfortunately non-rare) situation when the sampling is
unrepresentative of the population of allowed solutions, one may end
up with a wrong conclusion described by Xiao-Li Meng
\cite{MengXL:2018} as the {\it
  big-data paradox}, namely, ``the bigger the data, the surer we fool
ourselves.'' Sampling accuracy must be controlled in high-stake
phenomenological measurements, such as the recent measurement of $W$
boson mass by the CDF collaboration \cite{CDF:2022hxs}, together with
other uncertainties.

Incomplete sampling of PDF solutions may result in unstated sources of
the differences among central PDFs or the PDF uncertainties. The
possible existence of such differences is suggested by an observation
that, while several recent global analyses constrain the PDFs with
comparably strong sets of fitted experimental data, in some
phenomenologically important cases these analyses arrive at noticeably
different estimates of PDF uncertainties. In the CTEQ-TEA analyses,
some of the sampling uncertainty is included, as discussed below. On
the other hand, when the ``experimental" $1\sigma$ PDF uncertainty is
defined according to the $\Delta \chi^2=1$ criterion, as in ABM and
HERAPDF studies, this uncertainty does not account for sampling over a
sufficient class of PDF parametrizations and other factors, which must
be done separately.

As another example, the estimates for the correlated uncertainties on
key LHC and Tevatron total cross sections presented in
\cite{PDF4LHCWorkingGroup:2022cjn,Amoroso:2022eow} vary between the
recent PDF fits. In Fig.~\ref{fig:LHCTeVsigmaTot}, the 95\% confidence
level (C.L.) uncertainty regions on the $Z$, Higgs, and $W^\pm$ total
production cross sections at the LHC 14 TeV and Tevatron 1.96 TeV vary
in size in a large range. It has been demonstrated that the
uncertainties may reflect as much the fitting methodology as the
strength of experimental constraints.
 Indeed, while the differences with the non-global (ABMP'16 and ATLAS)
 and combined (PDF4LHC21) ensembles are reasonably understood, the
 differences between three global fits -- CT18, MSHT'20, and NNPDF3.1
 -- require additional attention.  When CT18 \cite{Hou:2019efy} and
 NNPDF3.1.1 \cite{Ball:2017nwa} NNLO PDFs were compared in Sec. 2 of
 the 2021 benchmarking study by the PDF4LHC group
 \cite{PDF4LHCWorkingGroup:2022cjn}, the former systematically
 predicted a larger uncertainty in the moderate $x$ region than the
 latter. The magnitudes of the MSHT20 NNLO uncertainties
 \cite{Bailey:2020ooq} in these comparisons tended to lie between the
 CT18 and NNPDF3.1 ones. More intriguingly, in the course of the
 PDF4LHC21 exercise, the three global PDF groups conducted fits to a
 set of common data, using common settings, so as to establish
 comparisons/benchmarks. The common data set (termed the ``reduced
 set'') was diverse enough to provide constraints on all PDF flavors,
 but limited enough so that all groups were expected to find similar
 estimates of PDF uncertainties. In the fits to the same reduced data
 set \cite[][Sec.~3, especially Figs. 3.4 and
   3.5]{PDF4LHCWorkingGroup:2022cjn}, the NNPDF3.1 (reduced)
 uncertainties came out to be systematically smaller than the CT18
 (reduced) and MSHT20 (reduced) uncertainties.

 The discrepancies in estimated uncertainties have a variety of
 implications. For example, they can explain different conclusions
 about the strength of evidence for the nonperturbative (intrinsic)
 charm component of the proton obtained by the NNPDF
 \cite{Ball:2022qks} and CTEQ-TEA \cite{Guzzi:2022rca} groups. They
 also affect projections for sensitivity of planned new experiments,
 just like related mathematical issues affect planning and policies in
 other fields \cite{Puy:2022a}.  Such differences are often attributed
 to the tolerance conventions chosen by the global analysis
 groups. ["Tolerance" refers to the prescription for estimating the
   PDF uncertainty, see the discussion in
   Ref.~\cite{Kovarik:2019xvh}.] Are the tolerance conventions mostly
 subjective, or can some conventions perform better than the others?
 The question is sharpened by formulating it as a problem about {\it
   sampling} of a {\it specific} PDF-dependent observable that PDFs
 themselves.

 Our article presents an introduction to the sampling of PDF
 solutions, followed by a presentation of a technique to improve
 sampling of PDF uncertainty for user-selected QCD observables.
 Section~\ref{sec:SamplingTrio} reviews mathematical essentials for
 this discussion. Among these, we first introduce the trio identity,
 useful for quantifying the convergence of sampling estimates. The
 trio identity for the sample deviation (Sec.~\ref{sec:trio}) and
 cornerstone properties of multidimensional (quasi-) Monte Carlo
 integration (Sec.~\ref{sec:QMC}) demonstrate that complex,
 large-scale analyses are at an elevated risk of an unaccounted
 sampling bias. Global QCD analyses must strive for representative
 sampling of all acceptable solutions, which may increase the
 resulting PDF uncertainties or effective tolerance.
 
 Section~\ref{sec:SamplingTrio} also points out fundamental
 difficulties in performing an all-inclusive test for representative
 sampling in a multi-parametric global fit. Such sweeping test is
 likely impractical. On the other hand, a practical question ``What is
 the sampling uncertainty on a given observable $X$?'' can be highly
 tractable using the already available technology for PDF fits. We
 point out the general rationale in Sec.~\ref{sec:QMC}. We then
 continue to Sec.~\ref{sec:hopscotch}, where we show that the question
 about PDF uncertainties on specific QCD observables can be explored
 using the general framework for large-scale surveys and QMC
 integration presented in Sec.~\ref{sec:SamplingTrio}.
 Section~\ref{sec:SamplingPDFs} reviews major types of sampling
 arising in PDF fits, from experimental data to models for
 systematics.  As a specific application, Sec.~\ref{sec:CT18NN40}
 investigates the PDF uncertainty on the LHC benchmark cross sections
 using the LHAPDF grids of NNPDF4.0 error sets and publicly available
 \texttt{mcgen}~\cite{Gao:2013bia,Hou:2016sho,MCGenwebsite} and
 \texttt{NNPDF}~\cite{NNPDF:2021uiq,NNPDF40Website} fitting codes to
 compute the $\chi^2$ of the included data sets. The hopscotch
 sampling technique introduced there suggests that the PDF uncertainty
 on key LHC cross sections at 13 TeV is larger than the nominal
 uncertainty obtained with the published NNPDF4.0 error sets.
 Section~\ref{sec:HopScotchTech} explains the algorithm of the
 hopscotch scans. Then, this section and Sec.~\ref{sec:t0Chi2} examine
 the PDF dependence of the log-likelihood for two commonly used models
 of experimental correlated systematic
 errors. Section~\ref{sec:HopscotchGood} offers a possible
 interpretation of our findings.  The PDF uncertainties must be also
 enlarged in the case of the strangeness-antistrangeness asymmetry and
 fitted charm PDF at large momentum fractions, as demonstrated in
 Sec.~\ref{sec:StrangenessAndCharm}.  Section~\ref{sec:Conclusions}
 contains conclusions.
 
\section{QCD sampling problem and the trio identity \label{sec:SamplingTrio}}

\subsection{Setup of the problem; the $R$ mechanism \label{sec:Setup}}
We start by discussing multi-dimensional sampling in a simplified
context, by considering the probability for a QCD observable $G$
dependent on the PDFs, such a collider cross section $\sigma$ or
perhaps the QCD coupling strength $\alpha_s$ determined from hadron
scattering measurements.  Predictions for observables are the ultimate
targets for the propagation of the PDF uncertainties.  The goal of the
physics endeavor is to estimate the truth value $G_\text{truth}$ of
$G$ that is objectively realized in Nature.  Historically, at most we
can hope to determine the expectation value $E_p(G) =
\frac{1}{N_p}\sum_{i=1}^{N_p} G_i$ on the population of many
measurements or other determinations $G_i$ of $G$, where $N_p$ is a
very large number of determinations.

We assume that the determinations $G_i$ are properly designed, so that
$E_p(G)$ agrees with $G_\text{truth}$ (i.e., $E_p(G)-G_\text{truth}$
is arbitrarily small) for $N_p$ that is sufficiently large. For
example, for $G=\alpha_s$, the population expectation $E_p(\alpha_s)$
could be computed on a future sample of many measurements obtained
after several more decades of well-funded research. If $G$ is a cross
section $\sigma$ computed with a multi-parameter PDF ensemble,
$E_p(\sigma)$ can be the expectation value with the PDF ensemble that
densely and representatively samples the whole parameter space.

The conundrum for many studies is that achieving such large $N_p$ may
not be feasible. Often one selects a sample of $N_s$ replicas from the
population, with $N_s < N_p$ or even $N_s \ll N_p$, and estimates the
sample expectation value as
\begin{equation}
E_s(G)= \frac{1}{N_s}\sum_{i=1}^{N_s} G_i = E_p(R
G)/E_p(R). \label{EsG}
\end{equation}
In the last step, we expressed the sample expectation $E_s(G)$ as a
ratio of population expectations $E_p(R G)$ and $E_p(R)$, where $R_i$
is an array of $N_p$ ``sampling indicators'' such that for each
element $G_i$ of the population
\begin{equation}
R_{i}=\left\{ \begin{matrix}1, & \mbox{if the }i\mbox{-th element is
    in the sample,}\\ 0, & \mbox{if it is not in the sample,}
\end{matrix}\right. \quad\quad \mbox{ for } i=1,\dots, N_p.
\label{Ri}
\end{equation}

The sample expectation deviation $\Delta_E \equiv E_s(G)-E_p(G)
\approx E_s(G) - G_\text{truth}$ is controlled by the accuracy of each
determination, or a replica in the case of PDFs, $G_i$, as well as by
the accuracy of {\it sampling} of $N_s$ replicas from the
population. The fitting accuracy/sampling accuracy distinction and the
representation using the $R$ indicators (``the $R$ mechanism'') are
borrowed from the study \cite{MengXL:2018} of large-scale surveys, in
which ``fits'' or ``replicas'' are equivalent to ``responses to the
survey''. Namely, the accuracy of a single replica $G_i$ can be raised
by reducing experimental, theoretical, and computational errors. From
here, we will assume that the individual $G_i$ are sufficiently
accurate. In contrast, the sampling accuracy reflects how adequately
we sample the population of $N_p$ acceptable replicas. If such
sampling is biased, the magnitude of the sample deviation can be
estimated using the $R$ mechanism, see Eq.~(\ref{trio}).

Small biases due to insufficiently representative sampling of large
populations may produce large deviations. Surveys of the COVID-19
vaccination rate with very large samples of responses and small
statistical uncertainties (e.g., Delphi-Facebook) greatly
overestimated the actual vaccination rate published by the Center for
Disease Control (CDC) after some time delay
\cite{BradleyMeng:2021a}. The deviation has been traced to the
sampling process. In contrast to the random error, which decreases as
$1/\sqrt{N_s}$, the sample expectation deviation can grow with both
$N_p$ and $N_s$.

Concurrently with the formalism for the large population surveys, a
related statistical formalism has been developed to understand
convergence of quasi-Monte Carlo (QMC) methods for multidimensional
integration \cite{Hickernell:2018a}. Insights from these formalisms
help us to elucidate our problem in the context of the PDF analysis,
in which it can be posed as follows:
\newtheorem*{problem}{Problem}
\begin{problem}
    Estimate an expectation value $E_{p}\left(G\right)$ of an
    observable $G$ on a {[}possibly unknown{]} population of $N_{p}$
    replicas, given a sample of $N_{s}$ values $G_i$, where
    $N_{s}<N_{p}$.
\end{problem}
To get such estimate, it suffices to adopt an $R$ mechanism that
renders $\Delta_E = E_s(G)-E_p(G) = 0$ within a prescribed error. In
this section, we discuss convergence of such sampling estimates.

\subsection{A toy example \label{sec:toy}}
\begin{figure}[b]
\centering \includegraphics[
  height=142pt,keepaspectratio]{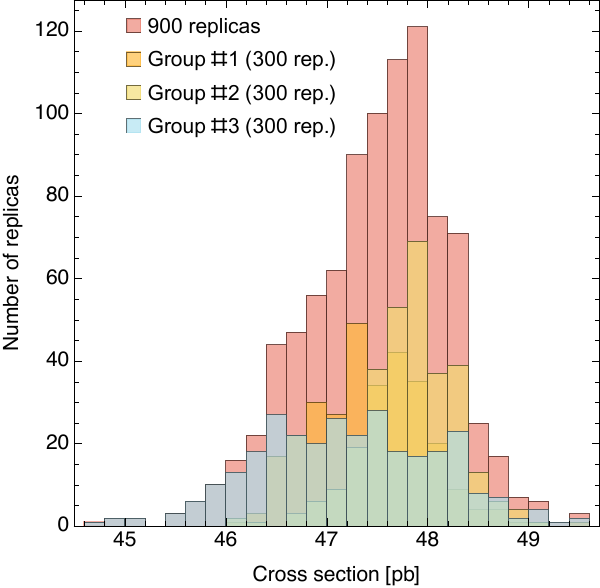}\quad
\includegraphics[
  height=142pt,keepaspectratio]{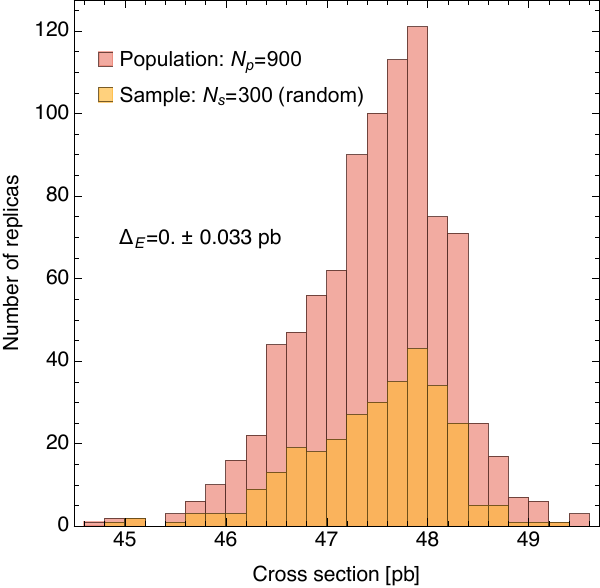}
  \quad
\includegraphics[
  height=142pt,keepaspectratio]{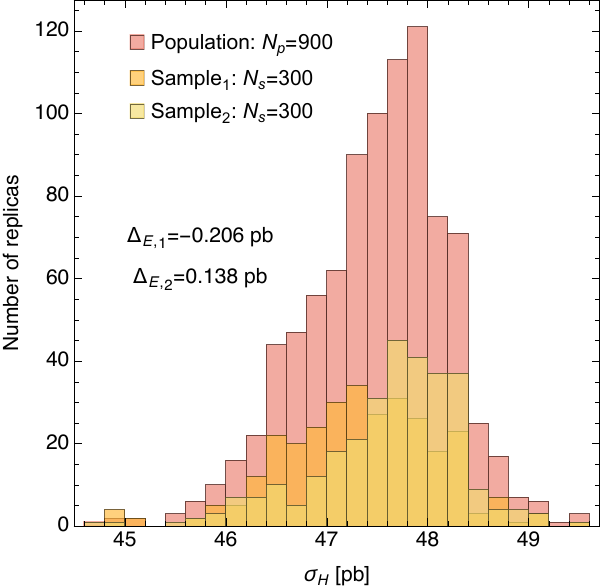}
\caption{ The $R$ mechanism illustrated on a toy example. The left
  panel shows a histogram of predictions for NNLO Higgs cross sections
  at the LHC based on 900 error PDFs of the PDF4LHC21 PDF ensemble
  (``the population"). The PDF4LHC21 ensemble is composed of three
  groups with 300 predictions; the resulting histograms are
  superimposed. The middle panel shows 300 predictions that are
  randomly sampled from the 900-member population, producing a sample
  average $\hat{\mu}$ that is indistinguishable from the population
  average $\mu$ (i.e., $\Delta_E \approx 0$). The right panel depicts
  two biased samples. Each contains 100 predictions from one group,
  200 from a second group, and none from the third. The biased
  samplings result in non-zero $\Delta_E$ values.}
\label{fig:Rmecha}
\end{figure}

As a toy example, consider a population of NNLO Higgs boson cross
sections ${G\equiv\sigma}_{gg\rightarrow H}$ at the LHC c.m. energy 14
TeV. The cross sections are computed with $N_p=900$ error sets of the
baseline PDF4LHC21 PDF ensemble \cite{PDF4LHCWorkingGroup:2022cjn}
consisting of 300 MSHT20, 300 NNPDF3.1.1, and 300 CT18$^\prime$
replicas, illustrated in the left panel of Fig.~\ref{fig:Rmecha}. [The
  replicas
  are ordered as in the actual 900-replica baseline ensemble. The mean
  cross section of the CT18$^\prime$ (NNPDF3.1.1) subset is slightly
  lower (higher) than the population mean.] We have $E_p(G)=47.492$ pb
and wish to obtain a close estimate by sampling only $N_s=300$
replicas out of 900.

If we randomly select $N_s=300$ replicas from the whole population, we
obtain $\Delta_E = 0 \pm 0.033$ pb, where the 68\% C.L. uncertainty is
computed by repeating the random selection 1000 times. In this case,
$E_s$ and $E_p$ are statistically indistinguishable (see the middle
panel of Fig.~\ref{fig:Rmecha}). It is known on general grounds that,
with the random sampling from the whole distribution, $\Delta_E$
decreases as $1/\sqrt{N_s}$, independently of $N_p$
\cite{MengXL:2018}.

As an instance of a different sampling, let us select 100 replicas
from each of the MSHT, NNPDF, and CT18 subsamples, for a total of
$N_s=300$ replicas. In this case, we still get $\Delta_E = 0\pm 0.03$
pb, i.e., no deviation. Since the PDF4LHC21 baseline set of 900
replicas is constructed by randomly selecting 300 replicas from each
of the MSHT20, NNPDF3.1.1, and CT18$^\prime$ 1000-replica samples, we
conclude that this PDF4LHC21 non-random combination prescription
introduces no appreciable deviation.

We can generate various sample combinations selecting 100 replicas
from one of the three groups, 200 replicas of second group, and none
of the third. Two of those are shown on the right panel of
Fig.~\ref{fig:Rmecha}, with deviations of $\Delta_E =-0.206\pm 0.036$
pb and $\Delta_E =0.138\pm 0.031$ pb.  In this instance, the bias was
introduced by hand, but in realistic situations the bias can arise
from apparently small departures from the random (probabilistic)
sampling at various stages of the analysis.

\subsection{The trio identity \label{sec:trio}}
The trio identity \cite{MengXL:2018,Hickernell:2018a} for the sample
expectation deviation $\Delta_E$ is a representation introduced to
examine convergence of the sampling algorithm. For our problem, the
trio identity takes the form
\begin{equation}
\Delta_E =
E_{s}\left(G\right)-E_{p}\left(G\right)=\frac{\mathrm{\text{Cov}}_{p}\left[R,G\right]}{E_{p}(R)}=\underbrace{\mathrm{\text{Corr}}_{p}\left[R,G\right]}_{\text{confounding~correlation}}\cdot\underbrace{\sqrt{\frac{N_{p}}{N_{s}}-1}}_{\mathrm{\text{measure~discrepancy}}}\cdot\underbrace{\mathrm{Var}_{p}\left[G\right]}_{\text{population~variation}}.
\label{trio}
\end{equation}
The three factors on the right-hand side are population expectations
with different dependencies on $N_s$ and $N_p$. In Eq.~(\ref{trio}),
$\mathrm{\text{Cov}}_{p}\left[A,B\right] \equiv E_p\left( (
A-E_p(A)\right) \left(B-E_p(B)) \right)$ is the population covariance.
{\it Variation} $\mathrm{Var}_{p}\left[G\right] \equiv
\sqrt{\mathrm{\text{Cov}}_{p}\left[G,G\right]}$ reflects the
complexity of the population distribution.\footnote{Variation
  $\mathrm{Var}_{p}\left[G\right]$ and standard deviation $\bar
  \sigma_G$ are related as $\bar \sigma_G^2 =
  N_p/(N_p-1)\ \mathrm{Var}_{p}\left[G\right]^2$.}
{\it Measure discrepancy} $\sqrt{N_p/N_s-1}$ is due to the mismatch
between the sizes of the population ($N_p$) and the sample
($N_s$). The {\it confounding correlation}
$\mathrm{\text{Corr}}_{p}\left[R,G\right]$ lies between -1 and 1. It
quantifies efficiency of the sampling algorithm in comparison to
simple random sampling. The confounding correlation reflects
methodology of the analysis.  Methodological correlations play a
central role in precise PDF analyses \cite{ Ball:2021dab}, together
with data-driven \cite{DataDrivenCorrelations} and theory-driven
\cite{NNPDF:2019vjt, NNPDF:2019ubu, Ball:2021icz, Forte:2020pyp}
correlations.

If the sampling exercise is repeated $N_R$ times while keeping the
same $N_s$, each time choosing a different $R$ array, one can estimate
a mean-square error (MSE) of the sample deviation for a given $R$
mechanism:
\begin{equation}
 \text{MSE}_{R}(\Delta_E) \equiv E_R (\Delta_E^2) = E_R
 (\text{Corr}_p\left[ R, G\right]^2) \cdot
 \left(\frac{N_p}{N_s}-1\right)\cdot \text{Va}r_p(G)^2.
\label{MSEDeltaE}
\end{equation}

The trio identity establishes dependence of the sample deviation
$\Delta_E$ on the sampling algorithm \cite{MengXL:2018}.
\begin{enumerate}
    \item Under simple random sampling (SRS), when replicas are
      independently selected with identical probability, the sample
      deviation converges to the truth as $1/\sqrt{N_s}$ in compliance
      with the law of large numbers:
    $$\mbox{SRS: } \Delta_E \to 0 \mbox{ as }N_s \to N_p,$$ with
    \begin{equation}
    \text{MSE}_\text{SRS}(\Delta_E) \equiv V_\text{SRS} = \kappa
    \ \bar \sigma_G^2/ N_s, \mbox{ where } \kappa\equiv (N_p -
    N_s)/N_p \sim 1\ .
    \label{MSESRSDeltaE}
    \end{equation}
    Comparison of Eqs.~(\ref{MSEDeltaE}) and (\ref{MSESRSDeltaE})
    shows that $E_\text{SRS} (\text{Corr}_p\left[ R, G\right]^2) =
    1/(N_p-1)$.
    \vspace{6pt}
    \item For an arbitrary sampling algorithm, the sample deviation
    satisfies
    \begin{eqnarray}
        &\Delta_E = \text{Corr}_p\left[ R, G\right] \sqrt{N_p-1}
        \sqrt{V_\text{SRS}}, \nonumber \\ &\text{MSE}_R(\Delta_E) =
        E_R (\text{Corr}_p\left[ R,
        G\right]^2)\ (N_p-1)\ V_\text{SRS}.
        \label{NoSRS}
    \end{eqnarray}
\end{enumerate}

For the sampling deviation to vanish as $N_s$ increases,
$\mathrm{\text{Corr}}_{p}\left[R,G\right]$ should decrease at least as
fast as $\text{o}(1/\sqrt{N_p-1})$. Absent this behavior,
unrepresentative sampling may lead to a situation when the sample
deviation remains large in spite of misleadingly small standard error
estimates. Meng dubbed this situation as ``the big-data paradox'',
which is clearly undesirable and unfortunately can go unnoticed if
sampling accuracy is not controlled to a sufficient degree.

\subsection{Quasi-Monte Carlo integration \label{sec:QMC}}
The trio identity elucidates why quasi-Monte-Carlo (QMC) methods for
multidimensional integration may converge at a faster or slower rate
compared to the Monte Carlo integration based on SRS
\cite{Hickernell:2018a}. When integration is performed over a unit
hypercube in $N_\text{par}$ dimensions, the sample deviation
$\Delta_E$ coincides with the {\it (hyper)cubature error} and can be
decomposed into three factors that play the same roles as in
Eq.~(\ref{trio}).

Of particular interest to us is the convergence of QMC integration
when $N_\text{par}$ is large.  In this limit, the minimal number of MC
replicas that guarantees a convergent integral for an arbitrary
integrand grows as $2^{N_{\text{par}}}$ \cite{Sloan:1997a}, reflecting
the curse of dimensionality that was pointed out long ago
\cite{Bellman:1961,Bishop:2006}.  Not only dense sampling of a
high-dimensional volume requires an exponentially growing number $N_p$
of replicas, such as $2^{N_\text{par}} \sim 10^{30}$ for
$N_\text{par}=100$; suppression of the confounding correlation to the
adequate degree is likely as a daunting feat.

The sample expectation of a QCD observable $G(\vec a)$ in PDF fits is
merely an integral of the weighted probability function
$P\left(\vec{a}\right)$ over $N_{\text{par}}$ PDF parameters
$\vec{a}$:
\begin{equation}
E_{s}\left(G\right)=\int
G\left(\vec{a}\right)\ P\left(\vec{a}\right)\ d\vec{a}\ .
\label{EsGint}
\end{equation}
We immediately conclude that convergence of $E_s\left(G\right)$ to the
truth for an arbitrary $G\left(\vec{a}\right)$ is not at all
guaranteed in a PDF fit that depends on too many parameters and does
not control for representative sampling.

In such a complex fit, one practically cannot know if the sample PDF
uncertainty covers the truth values for all $G(\vec{a})$.  On the
positive side, it follows from Eq.~(\ref{EsGint}) that, if $G(\vec a)$
is known to substantially depend only on a few components of
$\vec{a}$, estimation of $E_s(G)$ becomes highly tractable. The reason
is that the convergence rate of QMC integration is controlled by {\it
  the effective number of components}, i.e. directions in the
parameter space, along which the variance of the integrand is
significant \cite{Caflisch:1997ValuationOM}. If the number of such
components is small, integration can be arranged so as to give more
weight to the sampling of the manifold spanning the corresponding
``large dimensions''. For example, the coordinates in the subspace
with highest variances of $G(\vec{a})$ can be sampled most
densely. The coordinates in the complementary subspace with low
variances can be either fixed or sampled with a low
density. Techniques exist for ranking the $N_\text{par}$ coordinates
according to the variance of the integrand using the Analysis Of
Variance (ANOVA) \cite{LiuOwen:2006}, principal component analysis
(PCA), or another dimensionality reduction method. Accuracy of
integration can be iteratively improved by adding contributions from
the coordinates with lower variances \cite{Sloan:1998b}. See
discussion in Sec.~8 of Ref.~\cite{Hickernell:2018a}.

The role of effective dimensions in accounting for large uncertainties
from complex models, beyond Monte-Carlo integration, was recently
highlighted for the broader context of applied
science. Ref.~\cite{Puy:2022a} stresses the role of uncertainties in
the decision making of new policies in the real world, ``where
reliance on excessively complex and overconfident models may have
deleterious social-environmental consequences."

Experience with high-dimensional integration thus raises a warning for
the analyses that fit a large number of flexible functions using a
modest number of fitted replicas. While these analyses excel at
finding acceptable sets of functions describing the data, they are
nevertheless prone to the risk of a sampling bias that grows with the
dimensionality of the problem. Apparent reduction of the variance does
not eliminate this risk because of the big-data paradox quantified by
the trio identity. It has been known for a while that precise sampling
of $\chi^2$ in the vicinity of the global minimum becomes inefficient
with traditional MC replicas: the majority of such replicas have too
large $\Delta \chi^2$ because of high dimensionality of the parameter
space \cite[][Sec. 3.B]{Hou:2016sho}.  All-inclusive testing for
representative sampling thus is difficult with a lot of free
parameters. Fortunately, typical QCD cross sections depend on specific
combinations of PDFs that can be established using data set
diagonalization \cite{Pumplin:2009nm} (for example, implemented as
optimization of Hessian sets for specific experiments in the
\texttt{ePump} package \cite{Hou:2019gfw}) or a related
method. Sampling of a known PDF combination can be tested with a
greatly reduced cost based on the dimensionality-of-integration
argument presented above. Hopscotch scans described in the next
section realize such test in practice.

\section{Sampling tests and hopscotch scans \label{sec:hopscotch}}

\subsection{PDF uncertainties on QCD observables as a sampling problem}
\label{sec:SamplingPDFs}
Section~\ref{sec:SamplingTrio} summarized recent mathematical
approaches to statistical surveys of large data sets and QMC
integration of functions dependent on many parameters. In this
Section, we advance a viewpoint that the same approaches can guide
estimation of PDF dependence of specific QCD cross sections. In this
case, we consider a population of predictions $\{G_i\}$ for an
observable $G$ based on a large collection of PDF sets that will be
obtained in the future. Without the loss of generality, we assume that
the PDF sets are indexed by independent countable parameters and are
acceptable according to the goodness-of-fit criteria explained below.

A prediction based on one such PDF set plays the role of an individual
response to the survey, given by the numerical value
$G_i$. Predictions based on one published PDF ensemble can then be
viewed as a sample with the size $N_s$ that is smaller than the
population size $N_p$. Again without the loss of generality, we can
assume that the expectation values can be computed using the
unweighted average as in Eq.~(\ref{EsG}), or, if so necessary, using
the weighted average as in Ref.~\cite{MengXL:2018}.  The formalism
from survey studies \cite{MengXL:2018,BradleyMeng:2021a} then tells us
that, given the complexity of PDF models, confounding correlations may
dominate the sampling bias $\Delta_E$ even when the sample SRS
deviation, proportional to $1/\sqrt{N_s}$, is small.

Validation of representative sampling is thus as essential as the
tests of quality of individual fits, such as strong goodness-of-fit
tests on resulting PDFs \cite{Kovarik:2019xvh} and the closure test
\cite{Ball:2014uwa} of the agreement of a trial fit with a
predetermined truth value within the uncertainty. However, for an
all-out sampling test, the computation of the confounding correlation
in the trio identity, Eq.~(\ref{trio}), requires to know the
population distribution as an input, which is not known while the fits
are performed. The confounding correlation can be predicted to a
degree by using a model population distribution based on simulated
pseudodata in the same spirit as done in the closure test. On the
other hand, tests of representative sampling are simpler for QCD
observables with low effective dimensionality.

But what exactly is sampled in the PDF fits? Several types of PDF
sampling are performed based either on a known or unknown probability
distribution. The uncertainties from each sampling type may or may not
be included as a part of the final uncertainty. To illustrate how the
groups handle various types of sampling, we will compare two recent
NNLO PDF analyses, CT18 \cite{Hou:2019efy} based on the analytic
$\chi^2$ minimization and NNPDF4.0 based on the MC sampling of neural
network parametrizations of PDFs \cite{Ball:2021leu}. We outline some
common categories of sampling, leaving out technical details of
specific realizations.\footnote{In addition to the categories
discussed here, Monte-Carlo integration uncertainty of theoretical
calculations may be important in some cases.}

\paragraph{Sampling of experimental data sets} occurs when these data
sets are selected for the fit.  As a variation, only a part of the
data set can be included. Some data sets may be included with $\chi^2$
weights that are different from unity, as has been done in PDF fits
circa year 2000. If there are inconsistencies among the data sets,
inclusion of a data set from the global fit may result in a
larger-than-nominal shift of the expectation value. The associated
variation is latent in the PDF fits with significant tensions among
the experiments, including the recent global PDF analyses. The
strengths of tensions among the fitted PDF sets are comparable in
CT18, MSHT'20, and NNPDF3.1 fits, as reflected by $\chi^2/N_{\rm pt}$
values of experimental data sets in Tables 2.1-2.3 of
\cite{PDF4LHCWorkingGroup:2022cjn}. Such tensions can be identified
with techniques described in \cite{Kovarik:2019xvh,
  Collins:2001es}. Standard techniques for estimation of the
corresponding PDF uncertainty, like jackknife cross-validation
(computing the expectation value on an ensemble of fits with one
experiment left out at a time), are hardly practical in the global
fits. Instead, global PDF fits may resort to a remedy of increasing
the $\chi^2$ tolerance associated with one standard deviation from
$\Delta \chi^2=1$ to a larger value. More complex tolerance
prescriptions can be alternatively used \cite{Kovarik:2019xvh}. In the
CT18 family of PDFs, a special CT18Z ensemble is provided to obviate
the change in the PDFs upon the inclusion of the ATLAS 7 TeV $W/Z$
production \cite{Aaboud:2016btc} that runs into tension with dimuon
SIDIS experiments. The difference between the CT18 and CT18Z central
values can exceed the sum of 90\% intervals of two ensembles. The
MSHT'20 and NNPDF4.0 analyses publish only the PDF ensembles for the
default selection of experiments. Mutual consistency of the data sets
is thus a part of the data-quality requirement for the reduction of
uncertainties.  \\

 \paragraph{Sampling of experimental data fluctuations} is the most
 familiar type of sampling. The NNPDF and other (pre-)MC approaches
 generate PDF sets by resampling and cross-validation of the
 experimental data. In this paradigm, multiple replicas of the fitted
 data are constructed by randomly fluctuating the data's central
 values according to the experimental uncertainties. For each replica
 of data, the PDFs are found by fitting to the training part of the
 replica and simultaneously cross-validating against the
 complementary, control part. The final PDFs optimally agree with both
 training and control parts based on a criterion that depends on the
 log-likelihood function $\chi^2$ computed with respect to the
 fluctuated data. Expectation values are then computed using an
 unweighted average of predictions on an ensemble of such
 replicas. The NNPDF group \cite{Ball:2011gg} calls this approach
 ``importance sampling". It is called ``resampling" or ``bootstrap" by
 other groups.  

The CT approach, on the other hand, finds the best-fit PDFs by
minimization of the log-likelihood $\chi^2$ computed with respect to
unfluctuated data, {\it i.e.} the published data with specified
statistical and systematic errors (see below). Expectation values in
this approach are computed using the best-fit PDF, confidence
intervals are estimated using Hessian eigenvector (EV) sets. The CT
fit can also produce MC error sets, usually done by the conversion of
the final Hessian PDF sets \cite{Hou:2016sho}. Reciprocally, the
NNPDF4.0 MC replica sets have been also converted into an ensemble of
50 Hessian PDFs, which reproduces the expectations, standard
deviations, and correlations of the NNPDF4.0 MC replicas.  Resampling
of experimental uncertainties and conversions between Hessian and
Monte Carlo PDFs are well understood and numerically accurate. \\

 \paragraph{Sampling of PDF functional forms and fitting/training
   methodologies} is another common source of an explicit or latent
 uncertainty. It is independent from the data resampling
 uncertainty. In the discussion of data fluctuations, we assumed that
 the MC replicas are generated with the same training methodology,
 including the same choices for the sizes and contents of training and
 control partitions, as well as the same condition to finish replica
 training. We also assumed that the parametrization forms of the PDFs,
 whether given by an analytic function or by a neural network of a
 certain architecture, do not change in the course of an individual
 fit or training cycle. Such settings of the fit of course can also be
 varied.  

In contrast to the experimental uncertainty, these choices are
associated with the prior probability. One aspect of this kind is that
the final PDFs, whether produced by analytical minimization or an
AI/ML method, aim to describe the data without underfitting or
overfitting parts of data.  As a consequence of the fundamental
variance-bias dilemma \cite{Geman:1992a,Bishop:2006}, overfitting is
not sharply defined. Namely, a fit or ML training with an arbitrary
functional form can produce multiple solutions that balance between
agreeing with the (un)fluctuated data, having the fitted function with
high variation, and allowing for random noise. One therefore expects
some differences between the overfitting tests adopted by various
groups.

{\it Smoothness}, such as the absence of sharp features in acceptable
PDFs, is a related condition that does not necessarily imply data
overfitting. Both CT18 and NNPDF4.0 analyses require the PDFs to
satisfy conditions of smoothness, positivity, and integrability, again
according to varied prescriptions.

In the CT18 analysis, the candidate fits were repeated with more than
250 alternative functional forms and produced substantial spread of
PDF solutions. The tolerance of the published CT18 ensembles, such as
those shown in the next subsections, was increased so that their
Hessian PDF uncertainty covers the solutions obtained with the
alternative parametrizations. CT18 parametrizations utilize Bernstein
polynomials, which allow examination of a variety of flexible, yet
usually smooth, functional forms.

The NNPDF4.0 analysis adopts a specific optimized algorithm to select
the architecture, train neural networks, and impose smoothness and
other prior conditions. As a part of the algorithm, the final 100 or
1000 NNPDF4.0 replicas are selected from a larger pool of replicas,
many of which exhibit non-smooth, short-length features. The algorithm
is checked for self-consistency in a closure test by fitting idealized
pseudodata and verifying quantitative estimators such as the
bias-variance ratio and quantiles of $\Delta \chi^2$ for groups of
experiments.
The closure test demonstrates that the NNPDF4.0 optimized algorithm is
{\it sufficient} for generating well-behaving PDFs that agree with the
known "truth" PDFs.\footnote{Success of the closure test depends on
the targeted precision and accuracy of the truth-fit comparisons.}
Closure tests, however, do not prove that the use of the NNPDF4.0
settings is a {\it necessary condition} for obtaining well-behaving
solutions under acceptable variations in methodology. This especially
applies to the case of fitting inconsistent data sets. Other
algorithms, which vary in terms of hyperparameters, priors, and
similar setting choices, may produce PDF solutions that
enlarge the nominal uncertainties.

The availability of public NNPDF4.0 MC and Hessian PDFs, together with
the public NNPDF4.0 code, opens a possibility for a test to evaluate
performance of the NNPDF4.0 algorithm in finding the PDF
solutions. Consider the Bayesian likelihood-ratio test in the context
of PDF comparisons \cite{Kovarik:2019xvh,Soper:1994km}. Suppose two
PDF solutions, A and B, have the same likelihood, but solution A is
deemed unlikely compared to B based on the ratio of {\it posterior}
Bayesian probabilities. From this, we conclude that solution A is
disfavored because of its lower {\it prior} probability, not because
of its likelihood. Generalizing for a collection of QCD observables,
we can identify the regions populated by {\it new} predictions that
have the same or higher likelihood as the nominal NNPDF4.0
regions. Differences between these regions arise from the {\it prior}
conditions imposed on the new and nominal solutions. The ML universal
approximation theorem \cite{cybenko_approximation_1989,
  hornik_universal_1990, hornik_approximation_1991} implies that both
groups of solutions can be approximated by neural networks. The
proposed test therefore examines sampling over classes of eligible
functions or, in the ML language, eligible neural networks. If, in
addition to having low $\chi^2$ values, the new solutions pass all
other goodness-of-fit criteria, they must be accounted in the final
PDF uncertainty e.g. in the form of an enlarged tolerance. The design
and implementation of the test are explained in
Sec.~\ref{sec:HopScotchTech}.\\

\paragraph{Sampling of likelihood functions.} There is another
ambiguity to consider, associated with the approximation of the
likelihood in the PDF fits. Since the experiments rarely provide the
full likelihood, it is usually  expressed as $P(D|T)=\text{const}\cdot
\exp\left(-\chi^2(D,T)/2\right)$, where $\chi^2$ is constructed from
experimental data values $D_i$, theoretical predictions $T_i$, and
associated uncertainties. The log-likelihood $\chi^2$ enters the
figure-of-merit function in the fit, where it can be combined with
prior conditions or computed with respect to fluctuated $D_i$, as
done during the training of MC replicas. The log-likelihood $\chi^2$
is also used, not necessarily in the same form as during the fit, for
external comparisons of PDFs like the ones done in our
study. Non-Gaussianities of the errors are frequently neglected, and
various approximations are made to the correlated systematic errors,
which still lack full understanding \cite[Sec. 5 in
][]{Amoroso:2022eow}. These choices produce non-identical forms of
$\chi^2$ used by ATLAS \cite{ATLAS:2021vod}, HERA
\cite{Abramowicz:2015mha}, CT \cite{Hou:2019efy}, MSHT
\cite{Bailey:2020ooq}, and NNPDF \cite{Ball:2021leu}.  

In regard to the correlated errors, the PDF analyses address a common
ambiguity when converting percentage uncertainties into absolute
ones. For an experiment with $N_{\text{pt}}$ data points and
$N_\lambda$ systematic errors, the $\chi^2$ functions used by the
three global groups can be reduced to
\begin{equation}
    \chi^{2} = \sum_{i,j}^{N_{\text{pt}}} \left( T_{i} - D_{i}
    \right)\left(\text{cov}^{- 1} \right)_{ij}\left( T_{j} - D_{j}
    \right),
\end{equation}
where
\begin{equation}
  \left( \text{cov} \right)_{ij} = s_{i}^{2}\delta_{ij} + \sum_{\alpha
  = 1}^{N_{\lambda}}{\beta_{i,\alpha}\beta_{j,\alpha}}
\end{equation}
depends on uncorrelated uncertainties $s_i$, and correlated ones
$\beta_{i,\alpha}$. In turn,
\begin{equation}
\beta_{i,\alpha}=\sigma_{i,\alpha} X_i
\end{equation}
are derived from the tables of published $\sigma_{i,\alpha}$ using
unspecified normalization cross sections $X_i$. It has been observed
that plausible choices of $X_i$ nontrivially affect the resulting
PDFs. Search for the "least biasing" choices prompted scrupulous
investigations
\cite{Ball:2012wy,Pumplin:2002vw,Ball:2009qv,Gao:2013xoa}.  The
appendix in \cite{Ball:2012wy} reviews the rationales for these
choices, which depend on the type of the systematic error, while
Refs.~\cite{ATLAS:2021vod, Abramowicz:2015mha, Hou:2019efy,
  Bailey:2020ooq, Ball:2021leu} detail implementations of
$\beta_{i,\alpha}$ in the latest PDF fits.

The groups generally avoid fitting the PDFs with the choice of
$X_i=D_i$ for multiplicative errors (so called "experimental" scheme,
or "$exp$"), on the count that it was shown to bias the best-fit
results with respect to the truth in relatively simple examples
examined by D'Agostini \cite{DAgostini:1993arp, DAgostini:1999gfj} and
NNPDF \cite{Ball:2009qv}. Partly for this reason, CTEQ-TEA analyses
normalize all $\beta_{i,\alpha}$ by $X_i=T_i$ (the current theory)
\cite{Gao:2013xoa}. The NNPDF group uses a ``$t_0$ scheme", which has
been available in two versions: the pre-NNPDF3.0 analyses multiplied
only the normalization uncertainties by an iteratively updated theory
value, $X_i=T^{(0)}_{i}$ \cite{Ball:2009qv,Ball:2012wy}, and the
NNPDF3.0 and later analyses normalize all $\beta_{i,\alpha}$ with
$T^{(0)}_{i}$ in several groups of experiments
\cite[][Sec. 2.4.2]{Ball:2014uwa}, while the rest of the errors are
normalized by $X_i=D_i$.\footnote{In particular, the gluon at $x>0.1$
varies depending on the additive or multiplicative treatment of the
errors in inclusive jet production. For illustrations, see Fig.~18 in
\cite{Gao:2013xoa} and Figs. 60 and 61 in \cite{Ball:2014uwa}.}

These are not the only $\chi^2$ forms in use, however, and in fact the
NNPDF4.0 publication quantifies the quality of the fit and agreement
with the experiments with $\chi^2$ values in the $``exp"$ scheme
\cite[Sec.~5.1 in ][]{Ball:2021leu}. It can be understood that neither
of these conventions is safe from biases by recognizing that $X_i$ are
values of an initially unknown function $X$ that is fitted or learned
together with the PDFs. As such, $X$ is subject to the already
mentioned tradeoff between variance, bias, and noise
\cite{Geman:1992a,Bishop:2006}, with none of the current
implementations systematically controlling for this tradeoff. The
$``exp"$ and $t_0$ schemes correspond to the zero-bias (with respect
to $D_i$) and low-variance options, respectively, and a sequence of
other possible schemes lies in-between. [In the $``exp"$ scheme, the
  function $X$ goes through the fluctuating data points $D_i$. Other
  schemes use a smoother function.] The well-known demonstrations of
D'Agostini's bias assumed at most a few multiplicative errors. The PDF
fits deal with many multiplicative errors, whose pulls on the PDFs may
have opposite signs or be nonlinear, adding up to an unpredictable
effect. For high-statistics data samples, it is even possible that
random fluctuations in $D_i$ are smaller than uncertainties in
choosing $X_i=T^{(0)}_{i}$; and, finally, the truth $X_i$ for some
$\beta_{i,\alpha}$ in the experimental publication may not exactly
coincide with $T_{i}$ or its user-selected analog $T^{(0)}_{i}$ in the
PDF fit at hand.

The sampling test proposed above can also explore dependence on the
form of the likelihood, given that the NNPDF4.0 fitting code can
return $\chi^2$ values in the $``exp"$ and $t_0$ schemes.  As stated
in a note of the NNPDF4.0 code manual, ``the $t_0$ method is
\textbf{not} used by default by [applications of the
  \texttt{validphys} code other than replica training], and instead
the default is to compute the experimental $\chi^2$''
\cite{NNPDF40Manual}.  The sampling test can explore both prescriptions
as limiting cases for a family of schemes designed to estimate the
missing information in the provided correlation matrices.  The test
is agnostic about the generation of PDFs and
 just compares available PDF solutions without actually fitting them.

\subsection{CT18 and NNPDF4.0 probability regions for the LHC benchmark cross sections}
\label{sec:CT18NN40}
\begin{figure}[tb]
\centering
\includegraphics[width=.8\textwidth]{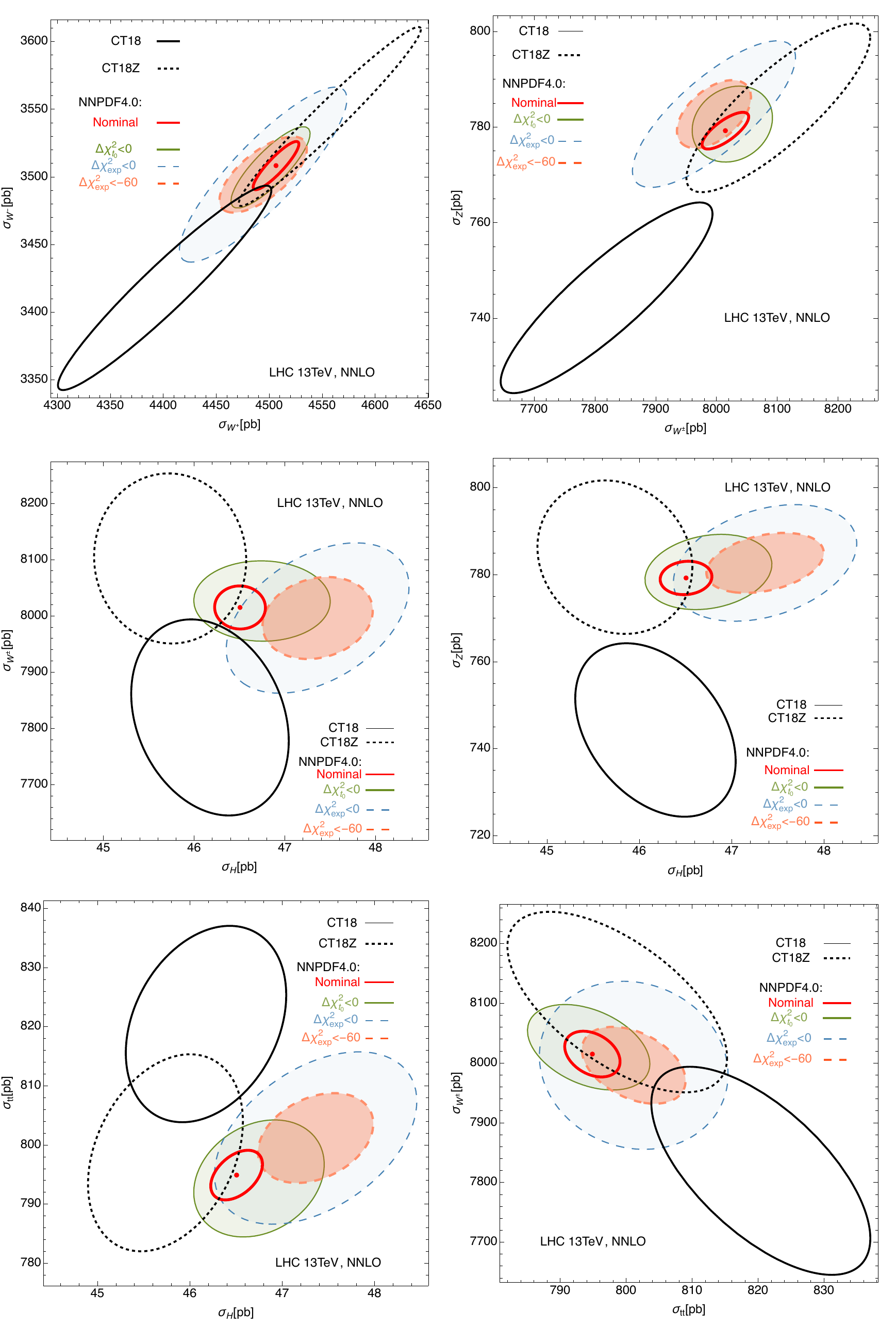}
\caption{ Black solid (dashed) ellipses: 68\% probability regions for
  select LHC cross sections at 13 TeV obtained with CT18 (CT18Z) NNLO
  PDFs \cite{Hou:2019efy}. Red solid ellipses: the nominal 68\%
  probability regions from the NNPDF4.0 NNLO analysis
  \cite{Ball:2021leu}.  Blue (brown) ellipses with a long-dashed
  contour: approximate regions with acceptable NNPDF4.0 solutions that
  have the same (better by 60 units) $\chi_{\mbox{\scriptsize exp}}^2$
  as the central PDF replica in the NNPDF4.0 publication. Green
  ellipses with a solid contour: same as blue, using the $t_0$
  prescription for $\chi^2$. The reconstruction of the approximate
  ellipses is described in Sec.~\ref{sec:HopScotchTech}.}
\label{fig:ellipses_1}
\end{figure}

The main findings of the test are summarized in
Fig.~\ref{fig:ellipses_1}, which shows the PDF uncertainties on LHC
cross sections at $\sqrt{s}=13$ TeV computed at NNLO in the QCD
coupling strength according to the settings listed in the
Appendix. For experimental collaborations, it is important to know
which theoretical predictions are acceptable given the latest
experimental and theoretical constraints. In this exercise, we
consider predictions based on the global fits that pass the
goodness-of-fit criteria adopted in the CT18 global QCD analysis
\cite{Hou:2019efy}.  All input PDFs used in the test are either error
PDFs available in the LHAPDF library \cite{LHAPDF6}, or linear
combinations thereof.

 The elliptical regions seen in Fig.~\ref{fig:ellipses_1} are
 projections of $N_{\rm eig}$-dimensional volumes populated by PDFs
 with low $\chi^2$ for the indicated Hessian PDF ensembles, where
 $N_{\rm eig}$ is the number of EV directions in the Hessian
 ensemble. For each Hessian PDF ensemble and $\chi^2$ scheme, there is
 one such approximately ellipsoidal volume. The ellipses of the same
 style in the six panels of Fig.~\ref{fig:ellipses_1} are the volume's
 projections onto the two-dimensional planes specified by the pairs of
 indicated total cross sections. {\it Here and in the following, the
   $\chi^2$ is computed with respect to the published (unfluctuated)
   central data values, except when we explicitly say otherwise.}

 Let us now go over the shown probability regions one-by-one. The
 black solid ellipses denote the 68\% C.L. regions obtained with CT18
 NNLO eigenvector PDFs in the analytical minimization framework
 \cite{Pumplin:2001ct}. The CT18 uncertainties are constructed so as
 to cover the solutions obtained in the CT18 fit using a large number
 of alternative parametrization forms and fit settings that were
 explored during preliminary CT18 fits. Thus, while the final CT18 PDF
 ensemble is provided with a single choice of the PDF functional
 forms, the CT18 PDF uncertainties reflect sampling over many
 (250-300) alternative parametrization forms, as well as variations in
 QCD scales of some experiments. The CT18 ensemble assumes $N_{\rm
   eig}=28$ EV directions.

In addition, we provide an alternative CT18Z PDF fit, in which the
strangeness and gluon PDFs are modified as a result of (a) including
the precision $W/Z$ production data set by ATLAS 7 TeV ($4.6\mbox{
  fb}^{-1}$) \cite{Aaboud:2016btc}, (b) using a factorization scale in
DIS that mimics small-$x$ saturation, and (c) other changes in the
selection of experiments and the charm mass. As with the CT18
ensemble, the nominal CT18Z uncertainty reflects solutions with the
alternative parametrization forms and settings. The CT18 and CT18Z
error bands are compatible at approximately 90\% probability
level. Confidence regions based on the CT18Z PDFs are shown as black
dashed ellipses. The shifts in the CT18Z predictions, as compared to
the CT18 ones, reflect to a large degree the inclusion of the ATLAS
$W/Z$ data set \cite{Aaboud:2016btc} which shows substantial tension
with the DIS experiments. Taken in the combination, the CT18 and CT18Z
confidence regions robustly predict the range of the outcomes based on
the various sampling options. The uncertainties obtained with this
prescription tend to be somewhat larger than the ones estimated using
the dynamic tolerance adopted by the MSHT group.\footnote{We have
verified this trend by comparing the PDF uncertainties in the CT18 fit
obtained using the CT and MSHT tolerance prescriptions.} On the other
hand, the dynamic tolerance estimates can be fragile if there are
large tensions among the experiments
\cite[][App.~A.4.b]{Hou:2019efy}. The CT approach is more robust to
such tensions.

It is interesting to compare the CT18(Z) uncertainties with those from
the NNPDF4.0 analysis \cite{Ball:2021leu} using either their
Monte-Carlo (MC) or Hessian error sets. Recall that CT and MSHT groups
perform analytic minimization of $\chi^2$ and provide Hessian EV sets
to estimate PDF uncertainties in applications. 

In the NNPDF4.0 approach, the Monte-Carlo error PDFs are constructed
by optimized training of neural networks on replicas of randomly
fluctuated experimental data.
 Each replica fit achieves a good $\chi^2$ with respect to its
 fluctuated data set, while practically all MC replicas have a very
 high $\chi^2$ (by hundreds of units) with respect to the unfluctuated
 data. The individual MC PDFs are thus poor fits to the published
 (unfluctuated) data set -- but their average (called the ``central
 replica", or ``replica 0") agrees with the unfluctuated data much
 better \cite{Hou:2016sho}.
The standard deviation on the ensemble of 1000 NNPDF replicas thus
essentially provides a 68\% experimental error with fixed
methodological settings, estimated around the central replica as the
spread of best fits to randomly fluctuated data points for the chosen
methodology, including selection (sampling) of experiments, training
and cross validation algorithm, and treatment of systematic effects.

 To examine the dependence of $\chi^2$ from the NNPDF4.0 code on PDF
 parameters, as discussed in Sec.~\ref{sec:SamplingPDFs}, we will
 employ the NNPDF4.0 Hessian EV set -- 50 error PDFs, with 1 error PDF
 per each of $N_{\rm eig}=50$ EV directions, obtained by post-fit
 conversion of NNPDF4.0 MC replicas. In the NNPDF implementation, the
 Hessian set is centered on the NNPDF4.0 replica 0 and reproduces
 $1\sigma$ symmetric PDF uncertainties of the MC set, i.e., the
 regions containing 68\% of the MC
 replicas. Section~\ref{sec:HopscotchGood} illustrates that, indeed,
 the NNPDF4.0 Hessian ensemble reproduces well these regions.

In the standard interpretation, however, Hessian eigenvectors form a
basis in linear space populated by vectors of PDF solutions in the
vicinity of the global minimum of $\chi^2$. The position of the global
minimum obviously depends on the $\chi^2$ scheme.  The approach that
provides 1 EV set per EV direction is usually based on the expectation
that the global minimum is very close to replica 0, and hence the
$\chi^2$ is more or less symmetric with respect to replica 0. We will
compare locations of replica 0 and global minima for two schemes, and
then examine the $\chi^2$ dependence in this space.

In Fig.~\ref{fig:ellipses_1}, red solid ellipses delineate the 68\%
probability regions computed with the published NNPDF4.0 error PDFs
and centered on the predictions from the NNPDF4.0 central
replica. These uncertainties are substantially smaller than the
CT18(Z) ones. 

 Overlayed on the nominal uncertainties, the other colored ellipses
 indicate {\it approximate} regions containing PDF solutions that have
 better $\chi^2$, according to the NNPDF fitting code, than the
 NNPDF4.0 central replica 0. Section~\ref{sec:SamplingPDFs} pointed
 out that comparisons in NNPDF publications use two forms of $\chi^2$
 as the figure-of-merit, called ``$t_0$" and ``$exp$" and computed
 with respect to the unfluctuated data, which differ in their
 implementation of experimental systematic uncertainties. The green
 and light blue ellipses delineate regions for each pair of cross
 sections in which our analysis have found regularly behaving PDF
 solutions that have $\Delta \chi^2 \equiv \chi^2-\chi^2_0 \lesssim 0$
 according to the ``$t_0$" and ``$exp$" definitions, respectively,
 where $\chi^2_0$ is computed for replica 0.  The brown ellipses are
 the analogous regions that contain the PDF solutions with $\Delta
 \chi^2 \lesssim -60$ according to the $``exp"$ definition. Inside the
 ellipses, $\chi^2$ shows quasi-Gaussian dependence on the PDF
 parameters with both definitions.  The found low-$\chi^2$ PDF
 solutions are linear superpositions of the NNPDF4.0 Hessian
 replicas. Among them, we find a few with $\Delta \chi^2$ as low as
 $-37$ ($-84$) with the $t_0$ ($exp$) definition.  Their $\chi^2$
 values are computed using the published NNPDF4.0 code
 \cite{NNPDF:2021uiq,NNPDF40Website}. We construct the alternative
 NNPDF4.0 solutions using an algorithm that we call a {\it hopscotch
   scan}, which performs focused sampling of PDF combinations giving
 the dominant contribution to the PDF uncertainty of the shown cross
 sections. Section~\ref{sec:HopScotchTech} explains this algorithm and
 construction of the approximate ellipses. The technique combines
 Lagrange Multiplier (LM) scans of PDF parameters \cite{Stump:2001gu}
 along the Hessian EV directions with stochastic sampling of the few
 ``large'' dimensions associated with the largest variance of the shown
 LHC cross sections, in accord with the general discussion in
 Sec.~\ref{sec:QMC}.

It is obvious from Fig.~\ref{fig:ellipses_1} that the ellipses
containing the PDFs that have $\Delta \chi^2 \leq 0$ are larger than
the nominal 68\% C.L. NNPDF4.0 uncertainties. These PDFs have been
examined for possible non-smooth features and other pathologies. We
did not observe obvious flaws and found them to be acceptable,
according to the CT18 procedure \cite{Hou:2019efy}, in light that they
achieve the same or better $\chi^2$ as the nominal fit, are smooth,
and fall within the nominal NNPDF4.0 errors nearly everywhere.

The two definitions of the $\chi^2$ reflect implementation of the
experimental correlated systematic errors. We pointed out that other
definitions exist, reflecting incomplete knowledge of systematic
uncertainties provided to the global fits \cite[][Sec. 5, and
  references therein]{Amoroso:2022eow}. Dependence on the $\chi^2$
definition must be scrutinized as a part of a more complete
exercise. Even within this limited scope, it is clear that there can
be many acceptable solutions with $\Delta \chi^2 \leq 0$ outside the
nominal NNPDF4.0 uncertainty. Section~\ref{sec:HopscotchGood} offers a
plausible interpretation of such alternative solutions. Our analysis
establishes the lower estimates on the corresponding regions in the
cross section planes, noting that the regions are comparable in size
to the CT18(Z) ones obtained with the two-tier tolerance and can be
shifted further from CT18 than the nominal NNPDF4.0 predictions.

\begin{figure}[b]
\centering
\includegraphics[width=.5\textwidth]{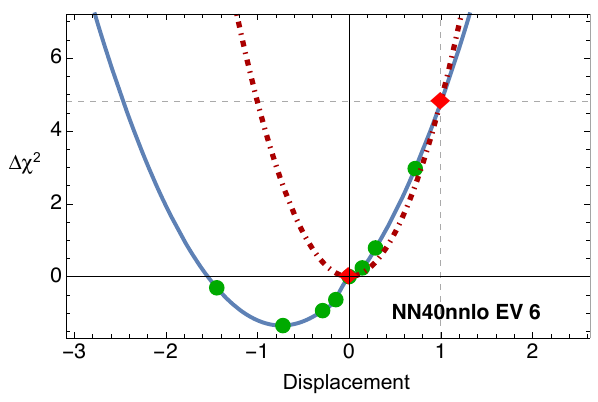}
\caption{Magnification of the $\chi^2$ scan for EV direction 6.  The
  green points and blue curve are the actual $\Delta \chi^2$ values
  and their interpolation from the scan. The dot-dashed red curve
  represents a symmetric parabola that would be obtained given only
  the central replica 0 and the EV set 6 (two red diamonds), and
  assuming that the third point necessary to build the parabola is the
  mirror of $\left(1\sigma,\Delta \chi^2(1\sigma)\right)$.}
\label{fig:LM_EV6}
\end{figure}

\begin{figure}[p]
\centering
\includegraphics[width=1\textwidth]{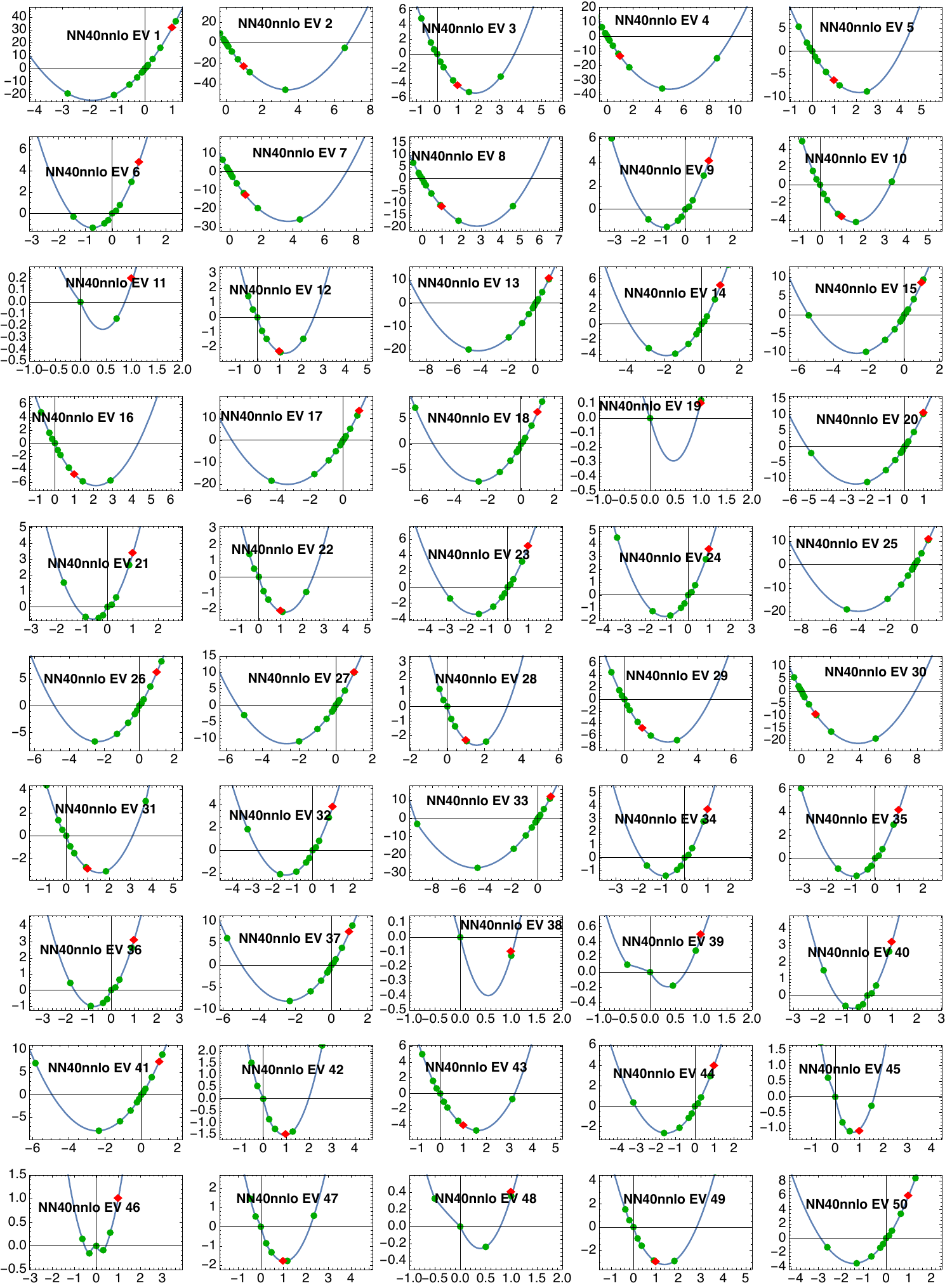}
\caption{Scans of the total $\chi^2$ in the experimental definition
  along 50 eigenvector directions of the NNPDF4.0 Hessian NNLO PDF
  ensemble. See text.}
\label{fig:parabolas_1}
\end{figure}

\subsection{A hopscotch scan, technical implementation}
\label{sec:HopScotchTech}

In this section we describe the construction of the alternative PDF
solutions that led to the ellipses of Fig.~\ref{fig:ellipses_1}. The
procedure realizes the general considerations in Sec.~\ref{sec:QMC},
namely:
\begin{enumerate}
    \item The NNPDF4.0 Hessian ensemble establishes natural basis
    coordinates $\vec a$ in space of MC replicas.
    \item For a typical QCD observable $G(\vec a)$, the largest
    variances are associated with 4-8 ``large dimensions'' in $\vec a$
    space.
    \item The PDF uncertainty on $G(\vec a)$ can be estimated with a
    moderate number of MC PDF replicas that vary along the large
    directions.
\end{enumerate}

We generate LHAPDF6 tables for the sample PDF replicas using the
\texttt{mcgen} program \cite{Gao:2013bia,Hou:2016sho,MCGenwebsite} and
the LHAPDF tables of the NNPDF4.0 NNLO Hessian ensemble as the
input. The total $\chi^2$ of the NNPDF4.0 analysis was evaluated using
the public code released by NNPDF~\cite{NNPDF:2021uiq,NNPDF40Website},
without refitting.  Specifically, the $\chi^2$ is computed by the
\texttt{perreplica\_chi2\_table} function of program
\texttt{validphys} included in the NNPDF code.  We activate the $t_0$
definition by setting option ``\texttt{use\_t0: True}" in the NNPDF
code and using the theory reference values for the
\texttt{210713-n3fit-001} PDF set provided with the NNPDF code.  The
kinematics cuts were fixed to be the same as in the NNPDF4.0 global
analysis \cite{Ball:2021leu}. The minimum values of $Q^2$ and $W^2$
for DIS measurements were hence chosen to be $3.49\mbox{ GeV}^2$ and
$12.5\mbox{ GeV}^2$, respectively.

The Hessian representation of the NNPDF4.0 ensemble provides the
central replica ($f_0$) and $N_{\rm eig}=50$ error PDF sets $f_i$
corresponding to displacements by a $+1\sigma$ value ($f_0+ \Delta
f_i$) along each independent eigenvector (EV) direction.  The total
$\Delta \chi^2_i$ of each EV set, computed with respect to replica 0,
varies among the individual EV sets, with some $\Delta \chi^2_i$ being
as large as $+35$ (for EV 1) or low as $-20$ (for EV 2) for the
experimental $\chi^2$ prescription, and with the majority no more than
$5-10$ units in magnitude. As only one error set is provided per EV
direction, this creates an expectation of an approximately symmetric
quadratic behavior of $\Delta \chi^2$ centered on $f_0$. This
expectation is illustrated in Fig.~\ref{fig:LM_EV6} for EV set 6 as a
red parabola, in which the red points correspond to replica 0 and EV
set 6. The horizontal axis is labeled in units of the $1\sigma$
displacement for EV set 6, and the vertical axis shows $\Delta
\chi^2$.

As an alternative to the red parabola, the actual $\Delta \chi^2$
behavior might have been very irregular, which may happen if NN fits
show large deviations from Gaussianity. To test which of the two
hypotheses is correct, we explicitly computed the $\Delta \chi^2$ at
green points, for which the LHAPDF tables are constructed as $f_0 +
w_i\, \Delta f_i$, where the real parameter $w_i$ quantifies the
displacement on the respective horizontal
axis. Figure~\ref{fig:parabolas_1} shows these $\Delta \chi^2$ scans
for all $N_{\rm eig}$ EV directions. In each EV direction, we evaluate
$\chi^2$ at 16 green points for a total of 800 points, with
Fig.~\ref{fig:parabolas_1} showing only the points with $\Delta
\chi^2$ below a few tens. We observe that $\Delta\chi^2$ follows
regular dependence consistent with a quadratic one along all EV
directions. However, the minima of the $\chi^2$ are displaced from the
central replica along many EV directions. Blue curves interpolating
the green points are consistent with symmetric parabolas whose minima,
$f_{i,\, min}\neq f_0$, are displaced from replica 0 in many EV
directions and render negative $\Delta \chi^2_{i,\, min}$ that can be
as low as $\approx -40$ (for EV 2). The widths of the reconstructed
parabolas vary noticeably. These observations strongly suggest the
regular, quasi-quadratic behavior of $\chi^2$ in the vicinity of the
central NNPDF4.0 replica and the existence of a displaced global
minimum in parameter space for which the $\chi^2$ is smaller than the
value provided by the central replica. EV sets and replicas with
negative $\Delta \chi^2$ were also pointed out in a thesis by the
NNPDF collaboration \cite{Lambri2020Milano}. Yet that study did not
provide further details, such as regular, approximately Gaussian
dependence of $\chi^2$ revealed by the hopscotch scans.

\begin{figure}[t]
\centering
\includegraphics[width=.48\textwidth]{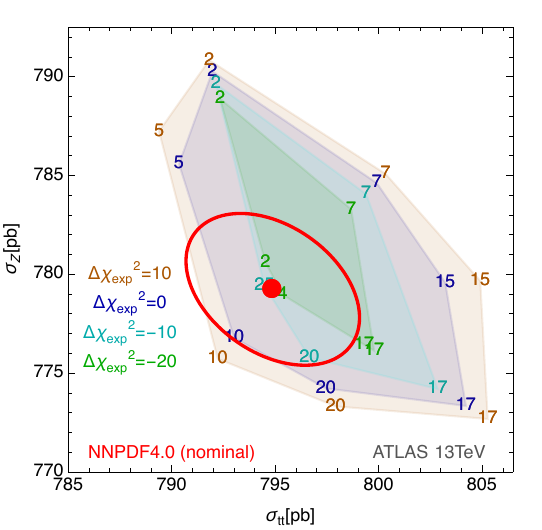}\quad
\includegraphics[width=.48\textwidth]{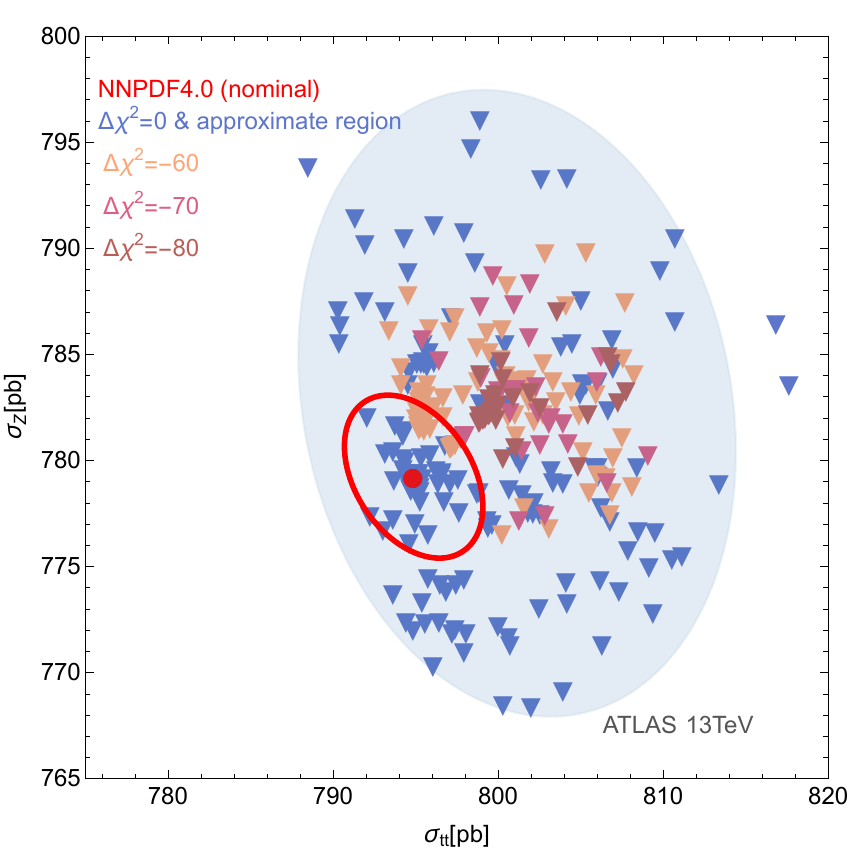}
\includegraphics[width=.48\textwidth]{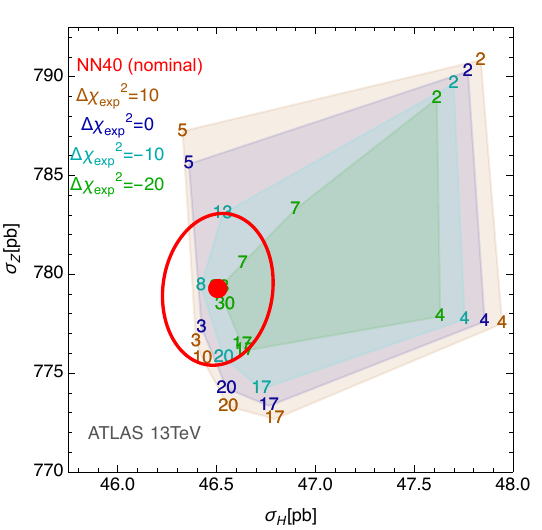}\quad
\includegraphics[width=.48\textwidth]{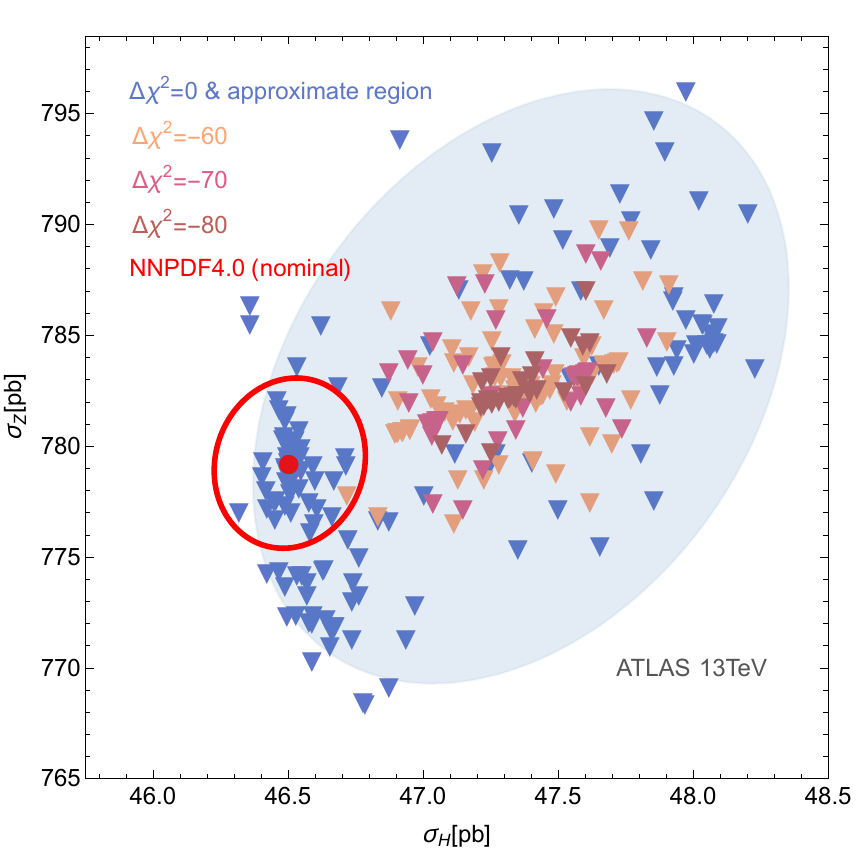}
\caption{Intermediate hopscotch scan results for $Z$ vs. $t\bar t$
  cross sections (upper row) and $Z$ vs. Higgs boson cross sections
  (lower row) for ATLAS at 13 TeV. See the Appendix for details of the
  computation. The left panels shows polygons formed by the pole sets
  with $\Delta \chi^2=+10,\, 0,\, -10$, and -20. In the right panels,
  the blue triangles correspond to $\Delta \chi^2=0$, with replicas
  with lower $\Delta\chi^2$ shown in increasing hue. Blue ellipses are
  approximate regions fitted to the $\Delta\chi^2<0$ boundary
  points. Red ellipses correspond to the 68\% probability regions from
  the published NNPDF4.0 Hessian set.  }
\label{fig:hopscotch_HZ}
\end{figure}

The hopscotch scan technique explores such low-$\chi^2$ region by
focusing on specific QCD cross sections. [Finding the displaced global
  minimum in the whole 50-dimensional space is more computationally
  expensive and beyond our study's scope, as complexity of
  combinatorial and geometrical factors increases quickly.] We draw a
low-dimensional ``court'' based on the $\chi^2$ behavior gleaned from
the EV direction scans and then repeatedly ``throw a marker''
according to one of the strategies to generate the PDF replicas at
points inside the court.

Initially, to find a region with replicas satisfying $\Delta \chi^2
\leq T^2$ in the plane of two cross sections, such as $\sigma_{t\bar
  t}$ and $\sigma_Z$, we use the interpolated parabolas in
Fig.~\ref{fig:parabolas_1} to find up to two ``pole'' PDF sets
corresponding to $\Delta \chi^2=T^2$ for each of 50 EV directions. We
plot the $\{\sigma_{t\bar t}, \sigma_Z\}$ pair for each pole set, as
is done for $T^2=+10,\, 0,\, -10,$ and $-20$ in the upper left panel
of Fig.~\ref{fig:hopscotch_HZ}. In the $N_{\rm eig}$ dimensional
space, the pole sets correspond to the corners of a rectangular block
whose projection on the $\{\sigma_{t\bar t}, \sigma_Z\}$ plane is a
polygon with the corners corresponding to the EV directions with the
largest displacements of cross sections from the central
predictions. In the upper left Fig.~\ref{fig:hopscotch_HZ}, these are
EV directions 5, 2, 7, 15, 17, 20, and 10. The other EV directions
(examined, but not shown in the figure) generate smaller
displacements. For this cross section pair, we generate $2\times 300$
replicas in the court consisting of two rectangular blocks spanning
the EV directions 5, 2, 7, 15, 17 and 17, 20, 6, 10, 5. A replica is
generated in an $n$-dimensional block as $f=f_0 + \sum_{i=1}^{n} w_i
\Delta f_i$, where each $w_i$ is a random real number that is
uniformly distributed along the $i$-th EV direction between the two
corresponding pole sets with $\Delta \chi^2 = T^2$.  We then generate
the replicas for three more pairs of cross sections: $Z$ vs $W^\pm$
(summed over the $W$ boson charges, EV directions 2, 7, 23, 20, 17,
5); $W^+$ vs $W^-$ (EV directions 2, 13, 1,17, 14); $t\bar t$ vs $H$
(EV directions 8, 15, 17, 4, 2, 5).

\begin{figure}[p]
\centering
\includegraphics[height=0.9\textheight]{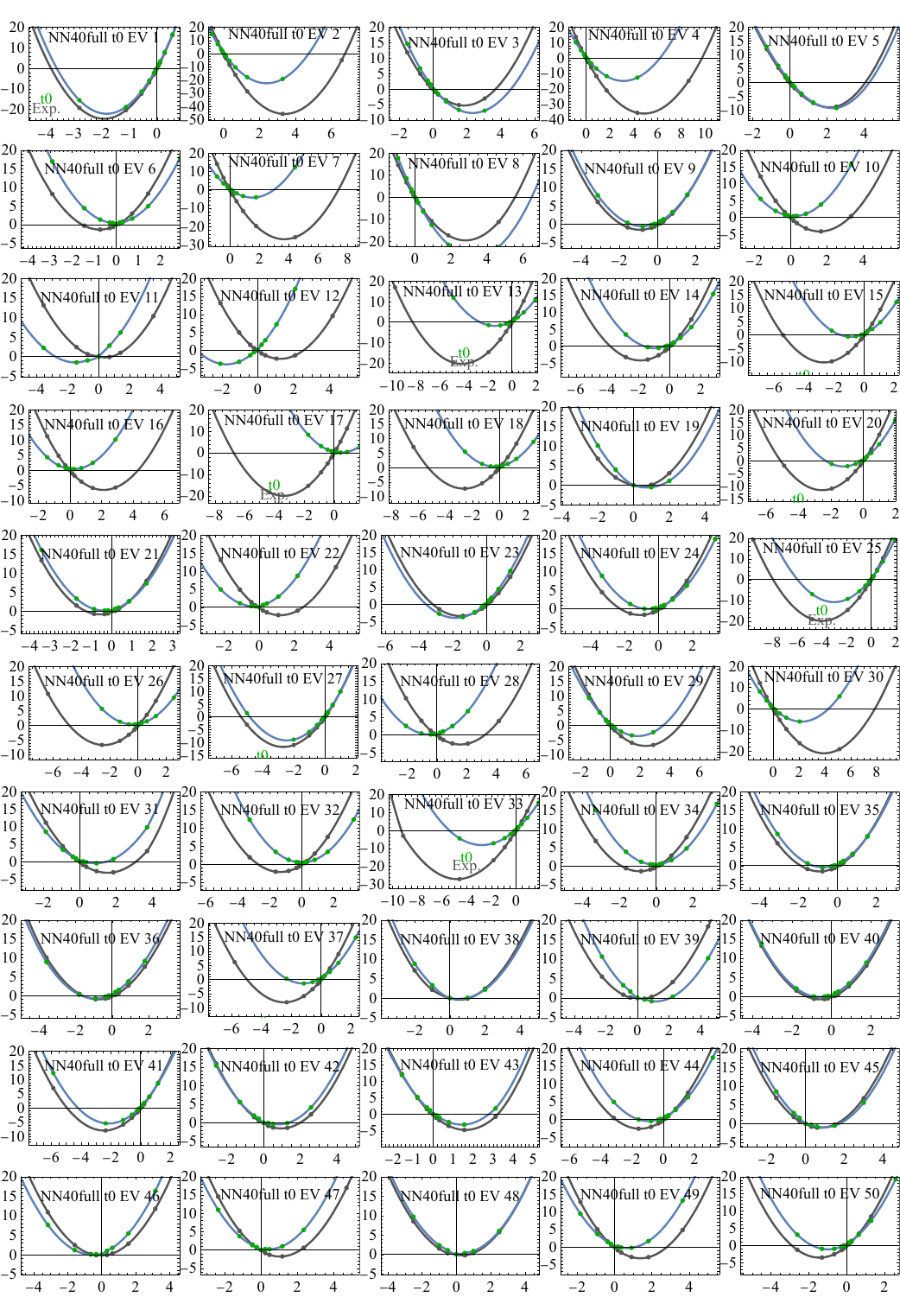}
\caption{Same as Fig.~\ref{fig:parabolas_1}, overlaying the $\chi^2$
  scans using the experimental ($``exp''$) and $t_0$ definitions of
  $\chi^2$.}
\label{fig:parabolas_2}
\end{figure}

Our cumulative set from all scans contains 2329 PDF ensembles to
examine solutions with $\Delta \chi^2 \leq 20$.\footnote{While we
refer to the hopscotch ensembles as ``replicas", they are not MC
replicas in the sense adopted in the NNPDF formalism. The hopscotch
replicas simultaneously have very good $\chi^2$ with respect to the
central data values and large displacements for the selected cross
sections. The traditional MC replicas are obtained by randomly
fluctuating the data or PDF parameters instead of directed search,
like the hopscotch scans: the majority of them have a positive $\Delta
\chi^2$ in the range of hundreds of units \cite{Hou:2016sho}.}
In the right column of Fig.~\ref{fig:hopscotch_HZ}, we use varied
colors to plot subsamples of replicas that have $\Delta\chi^2 \pm 3$
around the $\Delta \chi^2$ values specified in the figure. The
distribution of these replicas is consistent with the apparently
displaced global minimum, near which some replicas have $\Delta
\chi^2$ as low as -84 units. The lowest $\Delta \chi^2$ corresponds to
the regions populated by brown markers. In the lower row of
Figs.~\ref{fig:hopscotch_HZ}, we show the dominant EV directions and
replica samples for the $\sigma_Z$ vs. $\sigma_H$ pair, which was not
included in the generation of replicas. However, since this pair
shares the dominant directions with the sampled cross section pairs,
we can predict the PDF uncertainties for this pair as well.

 The hopscotch scan is mainly a search algorithm and, in the current
 realization, does not include any convergence criteria nor the
 certainty to find the true global minimum. [These aspects can be
 further developed along the lines discussed in Sec.~\ref{sec:QMC}.]
 The role of the hopscotch is to reduce the dimensionality of the
 search for solutions with a lower $\chi^2$ and to identify regions in
 the cross section space corresponding to such solutions.
 
While our set of solutions is not exhaustive, it can be used to
estimate the size of the projected area for a given value of $\Delta
\chi^2$, say $\Delta \chi^2 <0$. The sample's convex hull gives a
crude boundary of this region. On the other hand, since the EV scans
in Fig.~\ref{fig:parabolas_1} are strongly indicative of the
approximately Gaussian behavior of $\chi^2$, it seems reasonable to
assume that the populated regions in the cross section planes are
approximately elliptical.
With this information, a highly effective approach to estimate the
boundary is to fit an ellipse to the outermost points of the replica
subsample in the cross section plot. The quadratic form describing
each ellipse can be computed algebraically using a public Mathematica
program from \cite{Anwar:2019wrj} for reconstruction of
multi-dimensional ellipsoids from such projections. A 2-dimensional
ellipse can be reconstructed by having as few as 6 points on the
convex hull of the sample. In our case, we select no less than 15
outermost points per ellipse, so they can be fitted with good
certainty.

These approximate elliptical regions for $\Delta \chi^2<0$ are shown
in Figs.~\ref{fig:ellipses_1} and \ref{fig:hopscotch_HZ} in light
blue.  Given the distribution of hopscotch replicas for the
experimental $\chi^2$ prescription --see the right column of
Fig.~\ref{fig:hopscotch_HZ} -- the centers of the ellipses reside
close to the solutions with the minimal $\chi_{\mbox{\scriptsize
    exp}}^2$ found in our hopscotch scan. We also depicted the area
corresponding to $\Delta \chi_{\mbox{\scriptsize exp}}^2 < -60$ in
brown.

\subsection{Results for the $t_0$ definition of $\chi^2$}
\label{sec:t0Chi2}

As already noted, in the NNPDF4.0 article the PDF replicas are trained
using on the fluctuated training set using the $t_0$ definition and
additional conditions like positivity and integrability
\cite{NNPDF:2021uiq}. The post-fit comparisons employ both the
experimental ($``exp"$) and $t_0$ definitions with the full
unfluctuated data set. With the $t_0$ definition activated, the
total $\chi^2$ of replica 0 increases by about 340 units for $4618$
data points (from $\chi^2_{\rm tot}/N_{\rm pt}=$1.160 to 1.233)
compared to the $``exp"$ definition adopted in Sec.~5.1 of the
NNPDF4.0 publication to characterize the quality of the fit and review
agreement with the experiments.

Figure~\ref{fig:parabolas_2} compares the scans of the $\chi^2$ in the
experimental and $t_0$ definitions. We find that the $t_0$ definition
in Fig.~\ref{fig:parabolas_2} is also consistent with an approximately
quadratic behavior. Its minima along individual EV directions are
shallower and closer on average to the central replica than with the
experimental definition. Nevertheless, substantial shifts persist
along some EV directions, notably EV direction $1$ associated with the
small-$x$ gluon PDF.
In Fig.~\ref{fig:parabolas_2}, many hopscotch replicas with $-37 \leq
\Delta \chi^2 \leq 0$ with the $t_0$ definition still lie outside the
nominal NNPDF4.0 uncertainty, even though the difference from the
nominal uncertainty is smaller in this case than with the experimental
definition.  Repeating the hopscotch scans for the $t_0$
distributions, we establish the approximate ellipsoidal regions for
$\Delta\chi^2_{t_0} < 0$ in the planes of LHC cross sections shown in
green color in Fig.~\ref{fig:ellipses_1}.
In this case, the centers of the $t_0$ ellipses, being less shifted
than those obtained with the experimental prescription, are chosen to
be the centers of mass of the respective convex hulls.

\begin{figure}[tb]
\centering
\includegraphics[width=.45\textwidth]{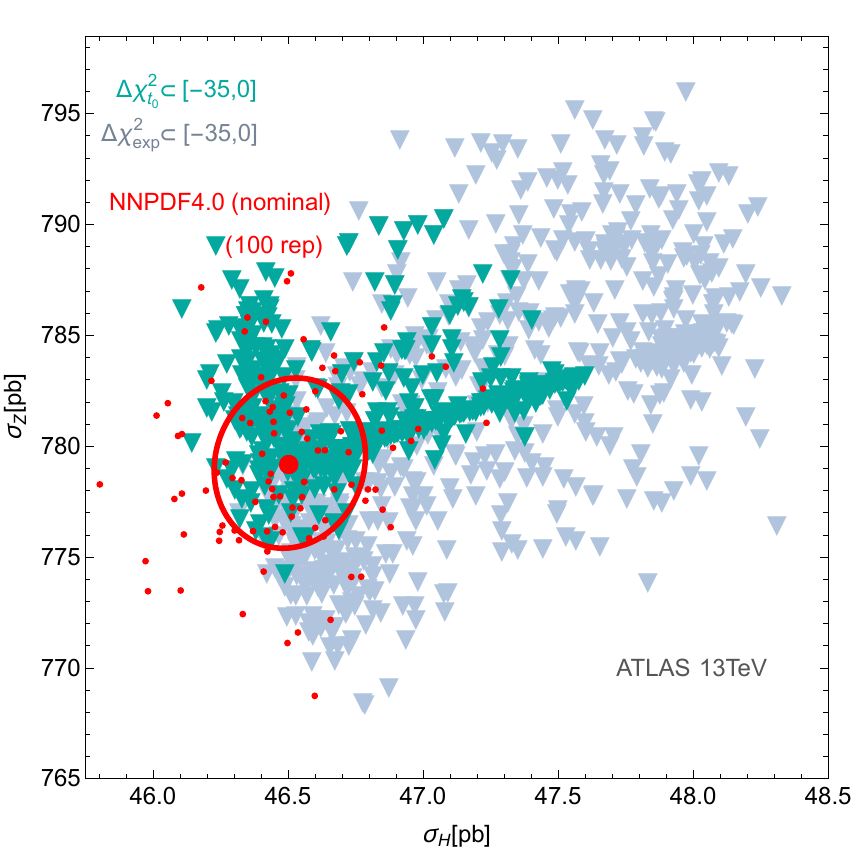}
\includegraphics[width=.45\textwidth]{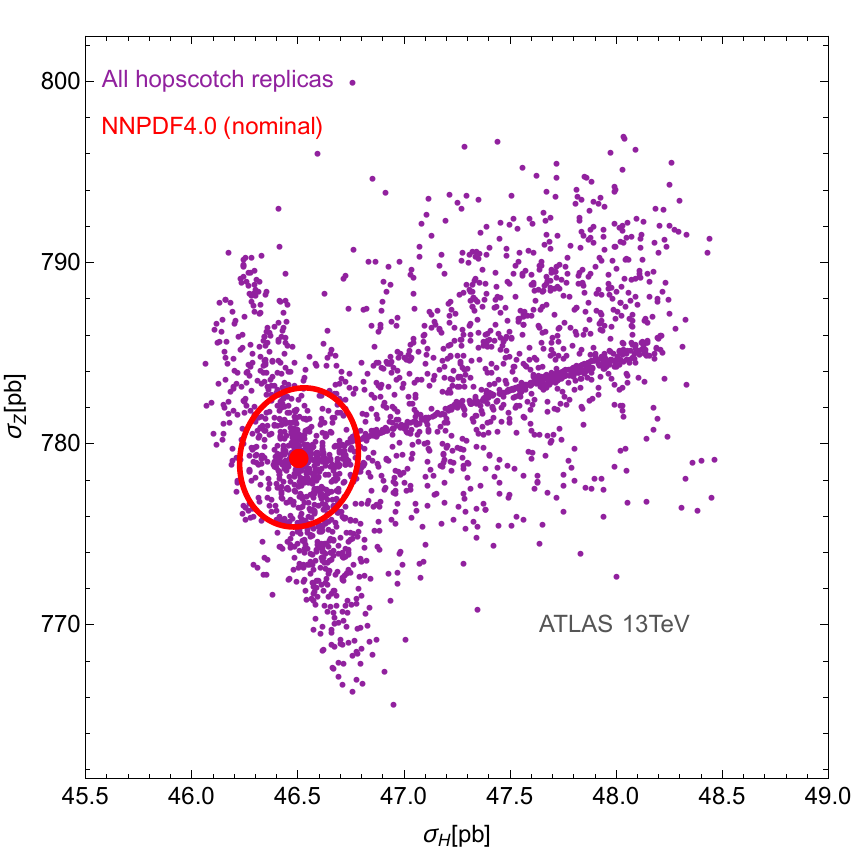}
\caption{Left: Hopscotch scan results for the Higgs vs. $Z$ cross
  section for ATLAS at 13 TeV. Here we show clouds of alternative
  replicas that have $-35 \leq \Delta \chi^2 \leq 0$ with respect to
  the NNPDF4.0 central replica, where $\chi^2$ is computed according
  to the $t_0$ (cyan) and experimental (grey) definitions. The red
  points indicate predictions with the 100-replica NNPDF4.0 ensemble.
  Right: The distribution of 2329 hopscotch replicas for any
  $\chi^2$.}
\label{fig:hopscotch_HZ_t0}
\end{figure}

\subsection{The hopscotch scans find the missing good solutions}
\label{sec:HopscotchGood}

The hopscotch exercise demonstrates the degree to which predictions
for LHC cross sections depend on the sampling procedures and priors
adopted by the groups. To the question: ``Which of our generated
replicas are acceptable for predicting the LHC cross sections?'', the
answer accounting only for the likelihoods is ``Apparently, all of
them that have good $\chi^2$'', echoing the likelihood-ratio test
described in Sec.~\ref{sec:SamplingPDFs}.

\begin{figure}[tb]
\centering
\includegraphics[width=0.49\textwidth]{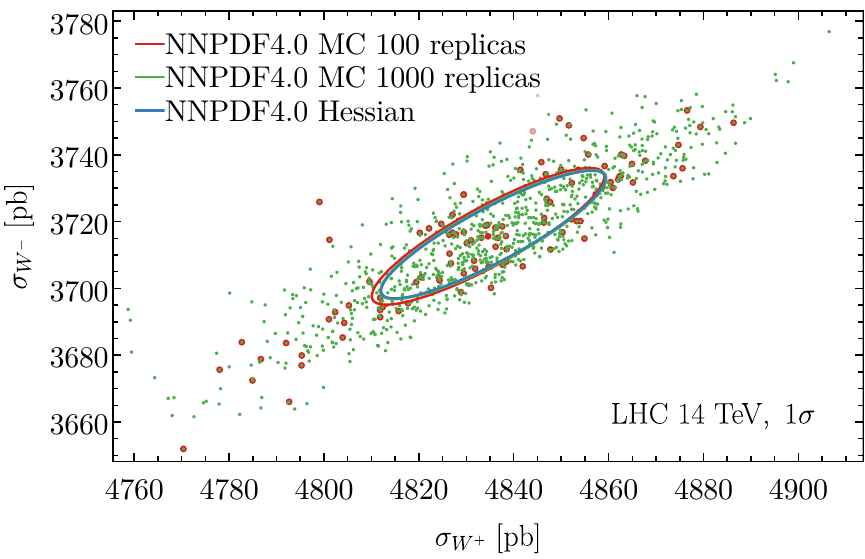}
\includegraphics[width=0.49\textwidth]{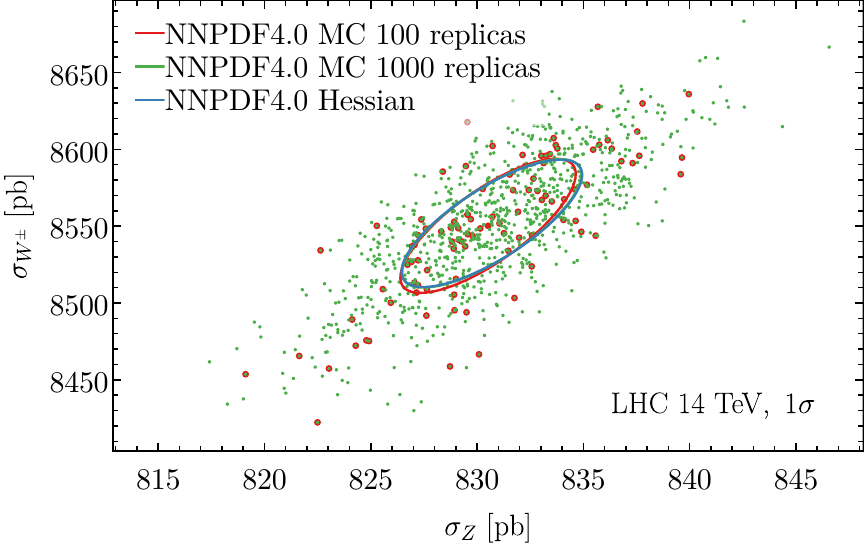}
\includegraphics[width=0.49\textwidth]{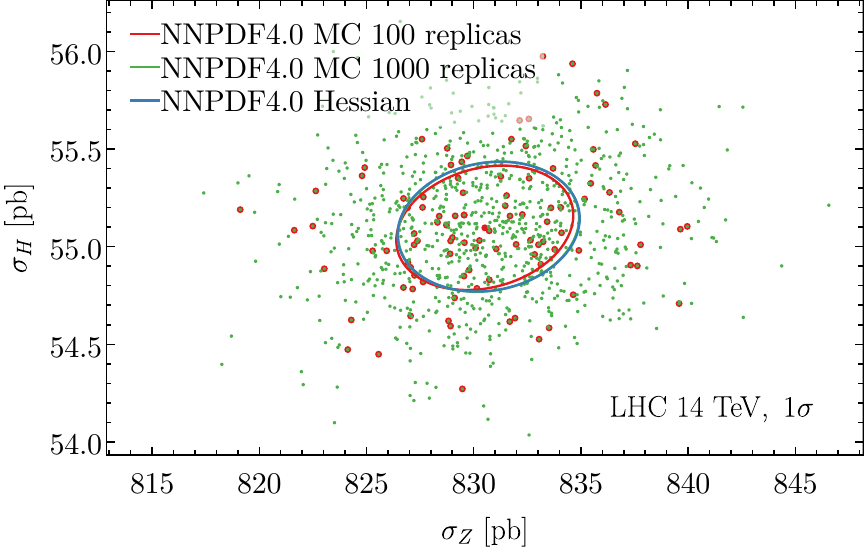}
\includegraphics[width=0.49\textwidth]{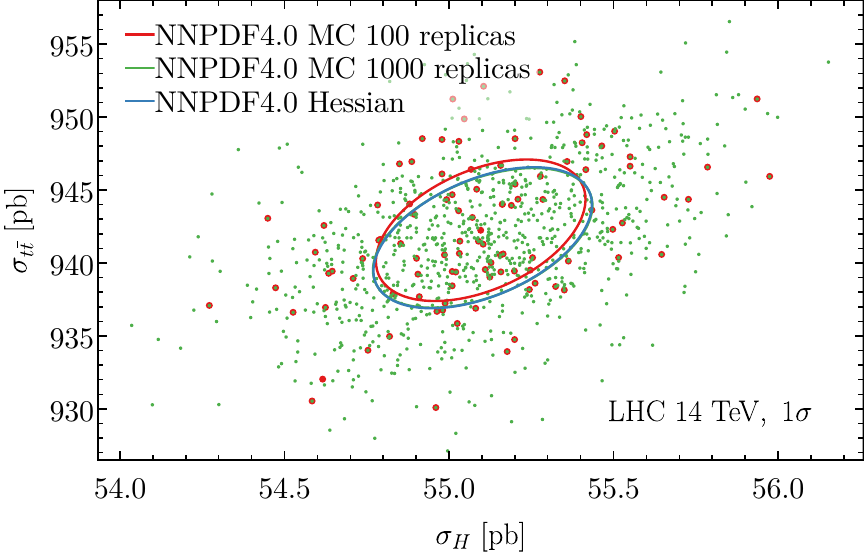}
\caption{LHC total cross sections at 14 TeV predicted using the
  NNPDF4.0 NNLO 1000-replica, 100-replica, and Hessian PDF
  ensembles. The ellipses indicate $1\sigma$ probability regions
  computed with each ensemble.}
\label{fig:NNPDFMCHessian}
\end{figure}

If we also want to explore the priors, seeking acceptable PDF
solutions becomes a notorious ``needle in a high-dimensional haystack"
issue recognized in studies of quasi-MC integration
\cite{Hickernell:2018a,Sloan:1997a,Sloan:1998b}.  To see this, let us
take a step back and recall that each NNPDF MC replica is specified by
a vector of a large size (of order 800 elements) containing NN latent
parameters. The closure test demonstrates that of order 1000 MC
replicas reproduce, within some accuracy, expected uncertainties in
the PDFs and predictions due to the fluctuations of the pseudodata
when training the replicas with a fixed methodology. When predicting a
vector of $N$ observables, predictions based on the MC replicas are
distributed relatively isotropically. This is illustrated in
Fig.~\ref{fig:hopscotch_HZ_t0}(left) and
Fig.~\ref{fig:NNPDFMCHessian}, where 2-dimensional projections of the
vectors of $N$ LHC cross sections, computed for the 100 nominal
replicas (red points) and 1000 replicas (green points), can be
converted into approximately spherical distributions by coordinate
rotations and rescalings.

Hessian PDFs provide a convenient eigenvector basis that captures PDF
variations in ``only'' 50 dominant dimensions around the NNPDF replica
0. Examinations of $\chi^2$ along the 50 EV directions in
Fig.~\ref{fig:parabolas_2} suggest that the global $\chi^2$ minimum is
displaced by many standard deviations with respect to replica 0 in a
direction that does not coincide with any EV direction. And, if we
identify a few EV directions that dominate a given cross section, we
can sample these directions more densely than allowed by the isotropic
sampling based on 1000 replicas.

Figure~\ref{fig:hopscotch_HZ_t0} illustrates how the hopscotch scans
perform targeted sampling of the parameter space based on the guidance
from the quasi-parabolic $\chi^2$ distributions in
Fig.~\ref{fig:parabolas_2}. For each selected pair of cross sections,
the hopscotch replicas densely populate a low-dimensional region in
the parameter space where $\chi^2$ decreases, while the cross sections
show high variability.
In Fig.~\ref{fig:hopscotch_HZ_t0}(left), we show predictions with the
hopscotch replicas that have $-35 < \Delta \chi^2 < 0$ with respect to
the $\chi^2$ of the NNPDF4.0 central replica, according to the chosen
$\chi^2$ definition. These regions of low $\Delta \chi^2$ are visibly
displaced with respect to the NNPDF4.0 central replica.
 The low-$\chi^2$ replicas are selected out of 2329 replicas that
 populate lower-dimensional hyperplanes in which $\chi^2$ decreases or
 increases slowly as a function of the selected cross sections. These
 hyperplanes and directions of the scans are identified based on
 Fig.~\ref{fig:parabolas_2}. In Fig.~\ref{fig:hopscotch_HZ_t0}(right),
 we see the projection of the distribution of 2329 replicas (with any
 $\chi^2$) on the $\sigma_Z$ vs $\sigma_H$ plane. The replicas are
 denser in the regions where the hyperplanes cross the projection
 plane. We remind the reader that this set of solutions is not
 exhaustive.
 
We already noted that both the NNPDF 100-replica ensemble and the
NNPDF Hessian ensemble reproduce well the underlying distribution of
replicas in their 1000-replica ensemble.
An illustration is provided in Fig.~\ref{fig:NNPDFMCHessian}, where
the LHC cross sections are predicted using the three ensembles. Here
the clouds of 100 replicas are consistent with the density
distributions of 1000 replicas. The 68\% probability regions given by
the ellipses are also consistent among the three ensembles.

The NNPDF4.0 Hessian ensemble employed for $\chi^2$ scans in
Fig.~\ref{fig:parabolas_2} captures overall properties of the
underlying replica distribution. Yet, the low density and distribution
of MC replicas does not capture the features of $\chi^2$ revealed by
the Hessian scans in Fig.~\ref{fig:parabolas_2} or predict the
parametric dependence of the replicas with the negative $\Delta
\chi^2$ that have been noticed before \cite{Lambri2020Milano}.

Upon a closer examination of the hopscotch scan, its generated
alternative PDFs for $\Delta \chi^2 \leq 0$ with either $\chi^2$
definition appear to pass the standard validation adopted in the CT
fits. They are linear combinations of well-behaving Hessian sets that
are sufficiently smooth and positive in the $x$ region with the data
constraints. At $Q=2$ GeV, only a few of them are negative in the
extrapolation regions, where their behavior can be easily adjusted
without changing the agreement with the data. We haven’t
scrutinized systematically the integrability of $T_3$ and $T_8$, as
done in the NNPDF4.0 fit, yet we observed no compelling reason to
discard these alternative solutions.

If the hopscotch solutions are acceptable, a natural question to raise
is why they are not covered by the nominal NNPDF4.0 ensemble. Since
these solutions have a good $\chi^2$, the conclusion from our test is
that they are disqualified by the NNPDF prior probability. Indeed, the
preliminary studies by NNPDF indicate that some of these replicas
(possibly a few dozen out of 2329) fail NNPDF4.0 requirements for
smoothness of PDF solutions \cite{UbialiHP2,Ball:2022uon}. If so, the dependence on
the priors would be best investigated in collaborative, comprehensive
benchmarking exercises among the PDF-fitting groups, using agreed-upon
criteria and computational tools.

We also observe that any hopscotch solution can be represented by a
neural network in accord with the universal approximation theorem
\cite{cybenko_approximation_1989, hornik_universal_1990,
  hornik_approximation_1991}. The challenge of representative sampling
in a high-dimensional space must therefore be also present in the NN
approach. We argued in Sec.~\ref{sec:SamplingPDFs} that the use of
data resampling (called "importance sampling" by NNPDF), combined with
a fixed methodology that makes specific choices for the NN
architecture, the cost function, stopping and smoothness conditions
\cite{Carrazza:2019mzf,NNPDF:2021uiq}, does not address samplings over
methodology-related settings at the various levels of the global
analysis. In the NNPDF4.0 analysis and closure test, the
hyperparameters of the methodology were optimized according to a
convention, not sampled in the optimum's vicinity. Variations in
training methodology are a part of the full uncertainty, together with
the theoretical uncertainty and another insufficiently understood
source of uncertainty due to the prescription for experimental
systematic errors. We have emphasized that the distribution of
replicas with good $\chi^2$ depends on the $\chi^2$ definition. This
dependence cannot be neglected at the contemporary accuracy level.

The discussion in this and earlier sections addresses the issues
raised in the NNPDF response \cite{Ball:2022uon} to our study. To
recap our relevant findings: 
\begin{enumerate}
  \item Random resampling of experimental data (``importance sampling'') does not account for variations in the training methodology. Alternative PDF solutions like those obtained by the hopscotch scans will be likely if the NN training methodology (i.e., the prior) is varied. 
\item  Most of the hopscotch solutions are sufficiently
  smooth upon a typical CTEQ-TEA examination and largely fall within
  NNPDF4.0 uncertainty bands, see the comparisons of PDFs at
  \cite{HopscotchWebsite}. It is unclear what fraction of 2330
  hopscotch solutions fails the smoothness test in
  \cite{Ball:2022uon}. Smoothness is not a sharply defined criterion, cf. the bias-variance dilemma.
\item  Arguments in favor of the $t_0$ prescription are based  on relatively
  simple examples. A variety of other
  $\chi^2$ prescriptions are in practical use. Representation of the
  systematic uncertainties with any available prescription is incomplete because of the
  bias-variance dilemma. NNPDF continues to use the experimental
  $\chi^2$ prescription for PDF comparisons in the NN4.0 publication
  and NN4.0 \texttt{validphys} code [except during NN training] \cite{NNPDF40Manual}. 
\end{enumerate}

\subsection{A case study: quark sea flavor composition and small-$x$ gluon}
\label{sec:StrangenessAndCharm}
\begin{figure}[tb]
\centering
\includegraphics[width=0.49\textwidth]{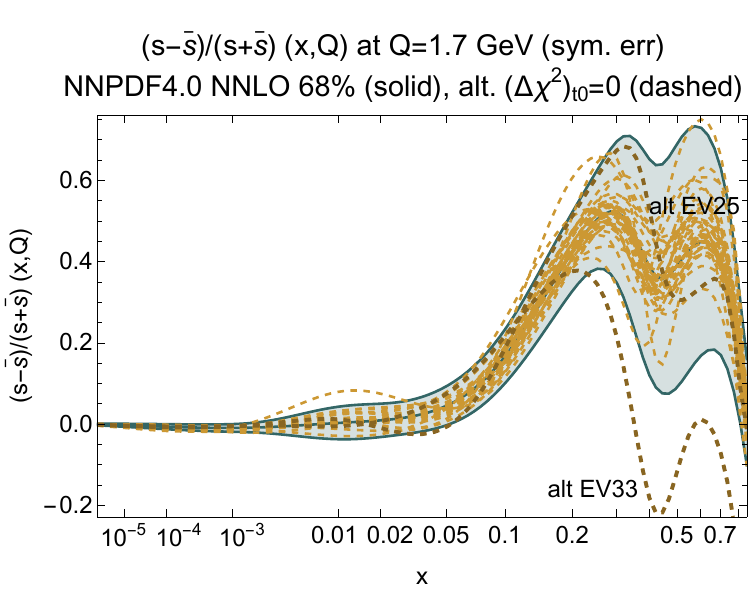}
\includegraphics[width=0.49\textwidth]{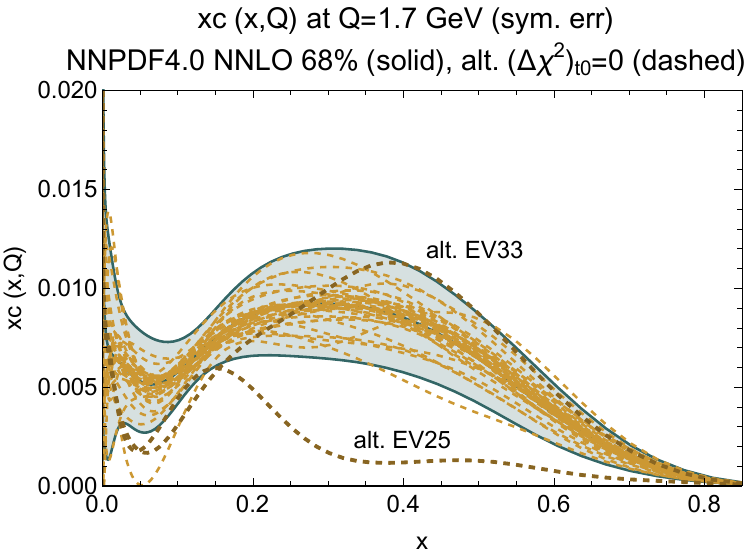}
\caption{Solid bands indicate the nominal 68\% NNPDF4.0 uncertainties
  for strangeness asymmetry (left) and charm PDF (right) at $Q=1.7$
  GeV. The alternative EV sets with $\Delta \chi^2_{t_0}=0$ are
  plotted as dashed lines. }
\label{fig:NNStrangeCharm}
\end{figure}

Implications of the hopscotch PDF solutions for uncertainties on
various QCD observables are of significant practical interest. The
$\chi^2$ scans along the NNPDF4.0 Hessian EV directions in
Fig.~\ref{fig:parabolas_2} indicate that, for each EV direction, there
is a displaced PDF set that has exactly the same $\chi^2_{t_0}$ (or
$\chi^2_{exp}$) as the NNPDF central replica 0. Just accounting for
these alternative sets can enlarge the nominal PDF uncertainty. On the
companion website \cite{HopscotchWebsite}, we provide the
\texttt{LHAPDF} grids for two 50-member ensembles of the alternative
sets with $\Delta \chi^2=0$ for the two $\chi^2$ definitions, as well
as figures comparing these PDFs with the nominal NNPDF4.0 NNLO
uncertainty bands.

For example, variations along Hessian EV directions 25 and 33
influence strongly the flavor composition of sea quarks and antiquarks
at $x>0.2$, where the relevant experimental constraints remain very
weak.  Figure~\ref{fig:NNStrangeCharm} presents two illustrations. The
left panel shows the nominal NNPDF4.0 uncertainty at the 68\%
probability for the strange-antistrange asymmetry,
$A_{\textrm{str}}(x,Q) \equiv \left(s(x,Q)-\bar
s(x,Q)\right)/\left(s(x,Q)+\bar s(x,Q)\right)$ at $Q=1.7$ GeV.
In the recent NNLO fits that allow strange quark and antiquark PDFs to
differ, the CT18As \cite{Hou:2022sdf,Hou:2022onq}, MSHT'20, and
NNPDF4.0 analyses all prefer a very large positive
$A_{\textrm{str}}(x,Q)$ at $x>0.3$, which can even exceed 100\% by
allowing the $\bar s$ PDF to go negative \cite[Sec. 4.5
  in][]{PDF4LHCWorkingGroup:2022cjn}. Such behavior may reflect some
tensions between the experiments. Among these fits, the positive
$A_{\textrm{str}}(x,Q)$ in NNPDF4.0 may be taken to be most
significant at $x\approx 0.2$, given the smallest nominal
uncertainty. However, the alternative EV set 33 for $\Delta
\chi^2_{t_0}=0$ in the left panel is consistent with a negative
$A_{\textrm{str}}(x,Q)$. From the plot of parabolas for EV direction
33 in Fig.~\ref{fig:parabolas_2}, we see that even deeper negative
variations of $A_{\textrm{str}}(x,Q)$ are allowed if $\chi^2_{exp}$ is
used, or if simultaneous variations along EV direction 33 and other EV
directions are considered. [Note that the EV directions specified by
  the NNPDF4.0 Hessian set do not change among the $\chi^2$
  definitions.]

The right panel of Fig.~\ref{fig:NNStrangeCharm} shows the counterpart
plot for charm PDF $c(x,Q)$ at $Q=1.7$ GeV. The nominal error band may
suggest a significant nominal enhancement of the charm PDF at
$x=0.2-0.3$, approximately in the same $x$ range where a large
$A_{\textrm{str}}(x,Q)$ appears in the left panel.
The hopscotch analysis shows that the uncertainty on charm PDF is
increased by considering the $\chi^2$ variations along the EV
directions revealed in Fig.~\ref{fig:parabolas_2}. Most notably, the
second $\Delta \chi^2_{t_0}=0$ set for EV direction 25 results in the
very small charm at $x>0.3$ at $Q=1.7$ GeV. When evolved down to
$Q<m_c =1.51$ GeV, this EV set will result in a vanishing fitted charm
at a low scale. After this set is included in the PDF uncertainty, the
NNPDF fit does not statistically prefer a nonzero fitted charm at the
initial scale, as would be concluded based on the nominal $1\sigma$
uncertainty \cite{Ball:2022qks}. Including variations along the other
EV directions, e.g., 33 that favors a smaller (larger) charm PDF at
$x=0.05-0.1$ ($x>0.4$), as well as uncertainties in the model for
systematic errors, further washes out the preference for the non-zero
fitted charm at large $x$ and low $Q$.  Indeed, the recent CT18 FC
analysis conveys that there is no evidence for intrinsic charm so
far~\cite{Guzzi:2022rca}.

Similar examinations for other PDF flavors and flavor combinations
(collected on the companion website \cite{HopscotchWebsite}) indicate
that the alternative $\Delta \chi^2=0$ solutions expand the
uncertainty on the gluon PDF at low $x$ and on the $T_3$ and $T_8$
combinations of quark and antiquark flavors.  One of the unexpected
findings of the NNPDF4.0 future test was that the fit without
including the HERA DIS data preferred the general growth of the gluon
PDF at $Q=1.65$ GeV and $x<10^{-3}$, where no constraints were
available. See Fig. 29 for $xg(x,Q)$ at $Q=1.65$ GeV in
\cite{Ball:2021leu}, where the solid green band for NNPDF4.0 without
the HERA data does not cover the blue and red bands that include these
data. A similar, also less pronounced trend is also seen with the
NNPDF3.1 methodology in their Fig. 28. Historically, the solutions
with the growing, flat, or even decreasing gluon at $x<10^{-2}$ were
allowed in the CTEQ fits in early 1990's, before the advent of the
HERA data.
Indeed, no experimental constraints existed in the pre-HERA data in
this region, similarly to the current situation with the nuclear PDFs
that have an essentially unconstrained gluon at $x<10^{-2}$ and may be
affected by strong nuclear shadowing.  See, for example, Fig.~6 in the
1995 ZEUS publication \cite{ZEUS:1995grj}, in which the pre-HERA data
at $W^2 < 500 \mbox{ GeV}^2$ and $Q^2 \geq 4.5\mbox{ GeV}^2$ do not
favor any particular trend of the $\gamma^* p$ total cross section at
$W^2 > 500 \mbox{ GeV}^2$, and hence they do not constrain the gluon
PDF at $x=Q^2/\left(W^2 + Q^2\right)\lesssim 0.01$ via scaling
violations.  Therefore, it is surprising that the NNPDF4.0 pre-HERA
future test disfavors the post-HERA small-$x$ gluon behavior.

The $\Delta \chi^2=0$ variations with the alternative EV sets 1, 2,
and 4 expand the nominal uncertainty in $xg(x,Q)$ of the full NNPDF4.0
set, especially in the downward direction at $x<10^{-2}$.  They modify
the NNPDF pre-HERA future tests, too.  In the same vein, considering
the hopscotch solutions indicates larger uncertainties on the flavor
combinations $T_3\equiv u+\bar u - d -\bar d$ and $T_8\equiv u+\bar u
+ d + \bar d -2 s - 2\bar s$ than seen in Fig. 49 of
Ref.~\cite{Ball:2021leu}.  We note that the hopscotch replicas agree
with the sum rules and integrability, especially as the PDF behaviors
at $x\to 0$ (outside of the data region) can always be adjusted to
obtain convergent first moments.

\section{Conclusions \label{sec:Conclusions}}
PDF uncertainties in high-stake measurements (Higgs cross sections,
$W$ boson mass…) should be examined for robustness of results to
sampling of available experimental data sets and PDF
parametrizations. Likewise, tests of manifestations of nonperturbative
QCD, such as the asymptotic large-$x$ behavior of intrinsic charm,
depend on interpretations of PDF uncertainties~\cite{Courtoy:2020fex,
  Ball:2022qks, Guzzi:2022rca}. Sampling biases may arise in PDF fits
operating with large populations of possible solutions. Increasing the
volumes of the fitted data and parametric space may increase, not
reduce, the sample expectation deviation. An undetected deviation may
result in a wrong prediction with a low nominal uncertainty.
 Sampling biases may limit reduction of the PDF uncertainties and
 explain some differences between the PDF sets.

For these reasons, global fits are potentially vulnerable to
unrepresentative sampling when their overall scope (including the
number of PDF parameters, size of data sets, range of possible
assumptions) grows. As a way to mitigate the risk of underestimation
in specific applications, statistical literature suggests to swap
democratic sampling in all dimensions for preferential sampling in
fewer dimensions that are most relevant to the task at hand.

In the Monte-Carlo (MC) replica method, constructing the Hessian
eigenvector (EV) sets from the MC PDF set introduces a convenient
coordinate system for such dimensionality reduction.  Taking the $W$
boson mass measurements as an example, we could identify the few
Hessian sets that give the largest contribution to the $M_W$ PDF
error. It is then more effective to sample these EV directions with a
higher density of replicas to look for acceptable PDFs that may be
outside of the nominal MC uncertainty. We presented a technique of
hopscotch scans to perform such estimation.

With this technique that does not require PDF refitting, we have
demonstrated that the NNPDF4.0 fitting code allows alternative
solutions of their global fit that predict the LHC cross sections
outside of the nominal NNPDF4.0 uncertainties, while having the same
total $\chi^2$ as the NNPDF4.0 central replica and satisfying typical
validation criteria adopted in the CT fit.
Literature on ML and analyses of high dimensionality suggests that
those solutions may exist and are not necessarily ruled out on the
basis of a low {\it prior} probability. Instead, for those solutions
that display an acceptable value of the likelihood, representative
sampling over methodological settings will contribute to the
confounding correlation. A related observation is that the dependence
of the distribution of acceptable predictions on the prescription for
implementation of experimental systematic errors cannot be neglected
at the targeted level of accuracy \cite[Sec. 5.1
  in][]{Amoroso:2022eow}. Here, we compare two $\chi^2$
 definitions -- ``experimental'' and ``$t_0$" -- readily available in
 the NNPDF4.0 code \cite{NNPDF40Manual},
 to probe how the likelihood-ratio test depends on
 the model of the likelihood [without doing the fit]. The regions
 populated by PDFs with the same $\chi^2$ according to the two
 definitions provide an upper estimate on the likely differences due
 to the choice of the $\chi^2$ prescription. In Sec.~\ref{sec:SamplingPDFs}, we argue
 that any prescription in use can generally produce a bias because of
 the variance-bias dilemma. We do not know the exact strength of the
 bias and emphasize that predictions based on the other definitions
 may fall in between the shown regions and should be further
 studied. Similar dependence has been observed in the CT fits (see
 e.g., Sec. 6D in \cite{Gao:2013xoa}).

 In the other two examples presented in Fig.~\ref{fig:NNStrangeCharm},
 we show that including the low-$\chi^2$ solutions from the hopscotch
 scans relaxes the NNPDF4.0 uncertainty on the flavor composition of
 sea quarks and antiquarks at $x>0.2$ and $Q<2$ GeV. As a result, both
 a negative strange-antistrange asymmetry and a zero fitted
 (intrinsic) charm PDF are statistically allowed at the $Q$ scale of
 order 1.5 GeV.

In either the MC or Hessian methods, a comprehensive range of fits
must be explored to understand variations due to the functional forms
and other choices.  This viewpoint is taken in the CTEQ-TEA family of
analyses, in which the tolerance on the fixed PDF functional form of
the published set is selected so as to cover candidate best-fit PDFs
found with the alternative choices. In other words, one must pay
attention both to the quality of accepted fits and their
representative sampling. For example, when some experiments disagree,
it should be either understood that fitting all experiments at once
will either fail the strong goodness-of-fit test
\cite{Kovarik:2019xvh} or, if such a fit is nevertheless accepted, the
tolerance may need to be increased, as the experimental tensions
suggest a larger uncertainty on the full population.

Instead of considering a large population of $N_p$ acceptable
solutions, for specific predictions, the trio identity equation
(\ref{trio}) can help to design a procedure that produces unbiased and
reliable estimates using a sample of a smaller size $N_s \ll N_p$. The
overall spirit of this approach is similar to data set diagonalization
\cite{Pumplin:2009nm} and replica unweighting
\cite{Ball:2011gg,Ball:2010gb}. The $R$ mechanism realizes a
generalization of such techniques and can select fits based on the
value of $\chi^2$ or other figures of merits.

We make LHAPDF6 grids of the alternative PDF replicas available for
the future analyses \cite{HopscotchWebsite}.

\acknowledgments We appreciate related investigations of CT18,
MSHT'20, and NNPDF3.1.1 Hessian and Monte-Carlo ensembles in the
context of PDF sensitivity studies with R. Thorne, E. Nocera, and
CTEQ-TEA and PDF4LHC21 working groups, as well as stimulating posts on
the sampling methods by S. Hossenfelder and A. Rojo Cruz. We are
grateful to Z. Kassabov for a technical communication on the NNPDF4.0
fitting code and to R. Thorne for a communication about the systematic
uncertainties.  AC is supported by the UNAM Grant No. DGAPA-PAPIIT
IN111222, German-Mexican research collaboration grant SP 778/4-1 (DFG)
and 278017 (CONACYT) and CONACyT – Ciencia de Frontera 2019
No. 51244 (FORDECYT-PRONACES).  PMN is partially supported by the
U.S. Department of Energy under Grant No. DE-SC0010129.  The work of
KX is supported by the U.S. Department of Energy under grant
No. DE-SC0007914, the U.S. National Science Foundation under Grant
No. PHY-2112829, and in part by the PITT PACC.  The work of CPY is
partially supported by the U.S. National Science Foundation under
Grant No. PHY- 2013791. CPY is also grateful for the support from the
Wu-Ki Tung endowed chair in particle physics.

\appendix

\section{Computation of hadronic cross sections}
\label{app:pheno}

In this section, we summarize settings of the computations of LHC
cross sections shown in the main part of the article. The cross
sections are computed at NNLO in the QCD coupling strength without
cuts, unless specified otherwise.

{\bf Drell-Yan $W^\pm/Z$ production}. For $W^\pm/Z$ boson production
at the Tevatron 1.96 TeV, we impose the CDF fiducial
cuts~\cite{CDF:2022hxs},
\begin{eqnarray}
W^{\pm}:& ~30<p_T^{\ell,\nu} < 55~\GeV, ~|\eta_{\ell}|<1,
~u_T<15~\GeV,~
60<m_T<100~\GeV;\\ Z:&~30<p_T^{\ell}<55~\GeV,~|\eta_{\ell}|<1,~u_T<15~\GeV,~
66<m_{\ell\bar{\ell}}<116~\GeV,
\end{eqnarray}
where
\begin{equation}
u_T=|\vec{p}_T^{\,\ell}+\vec{p}_T^{\,\nu(\bar{\ell})}|, ~
m_T=\sqrt{2(p_T^\ell
  p_T^\nu-\vec{p}_T^{\,\ell}\cdot \vec{p}_T^{\,\nu})}.
\end{equation}
For $W/Z$ boson production at the LHC, we adopt the ATLAS 13 TeV
fiducial cuts~\cite{ATLAS:2016fij},
\begin{eqnarray}
W^{\pm}:&~p_T^{\ell,\nu}>25~\GeV, ~|\eta_\ell|<2.5,~
m_{T}>50~\GeV;\\ Z:&~p_{T}^{\ell}>25~\GeV, ~ |\eta_\ell|<2.5, ~
66<m_{\ell\bar{\ell}}<116~\GeV.
\end{eqnarray}
The theoretical calculation is performed with a fast computation table
\texttt{APPLgrid}~\cite{APPLgrid} at NLO, combined with NNLO/NLO
point-by-point $K$-factors calculated with
\texttt{MCFM}~\cite{MCFM8,Campbell:2019dru}. The renormalization and
factorization scales are set equal to the invariant mass of the lepton
pair, $m_{\ell\bar \ell}$ or $m_{\ell\nu}$.

{\bf Top-quark pair production.}  Top-quark pair production is
measured by both ATLAS and CMS groups at 13
TeV~\cite{ATLAS:2016oxs,CMS:2018fks} and presented in the form of
total cross sections. Here we take the public code
\texttt{top++}~\cite{Top++} to compute these cross sections at NNLO,
with the threshold logarithms of soft gluons resummed up to the NNLL
level.  The factorization and renormalization scales are set to the
top-quark mass $m_t$.

{\bf Higgs production.} The calculation is done with
\texttt{ggHiggs}~\cite{Bonvini:2016frm} using the factorization and
renormalization scales equal to $m_H$.

{\bf Associated production of Higgs bosons and top-quark pairs.}
Recently a part of the NNLO calculation for $t\bar{t}H$ production
came out~\cite{Catani:2021cbl}, while no public code has been released
yet. Instead, we make predictions using
\texttt{MadGraph\_aMC@NLO}~\cite{Frederix:2018nkq} interfaced with
\texttt{PineAPPL}~\cite{Carrazza:2020gss} at NLO, and using NNLO PDFs.
The renormalization and factorization scales are set to be equal to
the partonic collision energy $\sqrt{\hat{s}}$.


\providecommand{\href}[2]{#2}\begingroup\raggedright\endgroup

\end{document}